\documentclass[twoside,twocolumn,english,aps,showpacs,prb,floatfix,reprint]{revtex4-2}
\usepackage{bm}
\usepackage{bbm} %fancy bold numbers
\usepackage{xfrac}
\usepackage{braket}
\usepackage{amsmath}
\usepackage{amssymb}
\usepackage{graphicx}
\DeclareGraphicsExtensions{.eps}

\usepackage{mwe}

\usepackage[
    bookmarks,
    bookmarksopen = true,
    bookmarksnumbered = true,
    pdfpagelabels,
    breaklinks = true,
    plainpages = false,
    linktocpage,
    colorlinks = true,
    linkcolor = blue,
    urlcolor  = blue,
    citecolor = blue,
    anchorcolor = green,
    hyperindex = true,
    hyperfigures
]{hyperref}
\usepackage{times}{\Large}
\usepackage[capitalise]{cleveref}

\DeclareMathOperator\arctanh{arctanh}
\newcommand{\avr}[1]{\langle #1\rangle}
\newcommand{\ve}[1]{\bm{#1}}
\newcommand{\intlim}[3]{\int\limits_{#2}^{#3}\!\!\mathrm{d}#1}
\newcommand{\wt}[1]{\widetilde{#1}}
\newcommand{\nn}{\nonumber}

\makeatother

\begin{document}

\title{Extended edge modes and disorder preservation of a symmetry-protected topological phase out of equilibrium}

\author{Thomas L.M. Lane}
\affiliation{Beijing Academy of Quantum Information Sciences, Beijing 100193, China}
\author{Mikl\'os Horv\'ath}
\affiliation{Beijing Academy of Quantum Information Sciences, Beijing 100193, China}
\author{Kristian Patrick}
\email[]{kpatrick@baqis.ac.cn}
\affiliation{Beijing Academy of Quantum Information Sciences, Beijing 100193, China}

\begin{abstract}
    The time evolution of topological systems is an active area of interest due to their expected applications in fault-tolerant quantum computing. Here, we analyze the dynamics of a noninteracting spinless fermion chain in its topological phase, quenched out of equilibrium by a Hamiltonian belonging to the same symmetry class. Due to particle-hole symmetry, the bulk properties of the system remain intact throughout its evolution. However, the boundary properties may be drastically altered, with the initially localized Majorana edge modes extending across the chain. Up to a timescale $t^*$, identified by area-law behavior of the entanglement entropy, these extended edge modes are an example of exotic effects in topological systems out of equilibrium. Further, while local disorder can be utilized to preserve localization and increase $t^*$, we still identify nontrivial dynamics in the Majorana polarization and Loschmidt echo.
\end{abstract}

\maketitle

%%%%%%%%%%%%%%%%%%%%%%%%%%%%%%%%%%%%%%%%%%%%%%%%%%%%%%%%%%%%%%%%%%%%%%%%%%%%%%%%%%%
%%%%%%%%%%%%%%%%%%%%%%%%%%%%%%%%%% INTRODUCTION %%%%%%%%%%%%%%%%%%%%%%%%%%%%%%%%%%%
%%%%%%%%%%%%%%%%%%%%%%%%%%%%%%%%%%%%%%%%%%%%%%%%%%%%%%%%%%%%%%%%%%%%%%%%%%%%%%%%%%%

\section{Introduction}

noninteracting symmetry-protected topological (SPT) phases of matter are classified by the tenfold way~\cite{altland_1997,kitaev_2009,ryu_2010,chiu_2016}: a classification scheme that uses discrete symmetries of the underlying Hamiltonian to quantify the potential topological phases. while this is well celebrated, it does not capture the possible phases of systems out of equilibrium. In this setting, due to the unitarity of the time evolution operator, only particle-hole symmetry (PHS) is preserved. Therefore, nonequilibrium systems require a different classification scheme than the static system, even if the Hamiltonians describing the initial system and the quench possess the same symmetries~\cite{mcginley_2019}. This can have striking consequences for SPT phases out of equilibrium, where the presence and nature of edge excitations may be dynamically modified~\cite{chung_2013,hegde_2015,hegde_2016,chung_2016,mcginley_2018}.

It has been shown in Ref.~\cite{mcginley_2019} that an out of equilibrium classification scheme, similar to that of the equilibrium setting, is established through a dimensional reduction procedure where topological properties of physical systems can be derived from higher dimensional systems of which they descend~\cite{qi_2008}. Importantly, the higher dimensional system must respect the same symmetries as the physical system. However, within this general framework, it is unclear whether the time-reversal (TRS) and chiral (ChS) symmetries are explicitly or dynamically broken. Regardless, it is only necessary to consider the existence (or not) of PHS. Consequently, all out of equilibrium systems will collapse onto an $A$, $C$ or $D$ classification.

Unfortunately, probes for dynamical changes in a topological phase are not trivially inherited from the analysis in equilibrium phases. As an example, consider a quench of a topologically nontrivial system defined by initial Hamiltonian $H^{(i)}$ and time evolution operator $U(t)=e^{-iH^{(f)}t}$, where $H^{(f)}$ is the postquench (final) Hamiltonian. The evolved system is described by the fictitious Hamiltonian $H_\mathrm{fic}(t)=U(t)H^{(i)}U^\dagger(t)$, such that the time evolved many-body state $\ket{\Psi(t)}=U(t)\ket{\Psi}$ is an eigenstate of $H_\mathrm{fic}(t)$. Since $H^{(i)}$ and $H_\mathrm{fic}(t)$ share the same, gapped spectrum, one can be continuously deformed into the other. Then, provided this is performed in a symmetry-preserving manner, they will also share the same bulk topological properties and, thus, a bulk topological invariant will no longer capture dynamical changes in $\ket{\Psi(t)}$. To resolve this, it has been shown that degeneracies in the many-body entanglement spectrum (ES), manifesting as zero modes in the single-particle ES~\cite{mcginley_2018,mcginley_2019}, may alternatively be used to identify dynamical changes of the topological phase~\cite{gong_2018,sedlmayr_2018,pastori_2020,sayyad_2021,zhang_2022,fang_2022,wu_2023}. Further, it is suggested that there exists a time $t^*\sim N/\nu_{LR}$, which scales with the size of the system $N$ and the Lieb-Robinson velocity $\nu_{LR}$, beyond which the topology of the system becomes ill-defined~\cite{mcginley_2019}.

In this paper, we expose nontrivial dynamics when $t<t^*$, which we use to identify a weak breakdown in bulk-boundary correspondence. We show that the inclusion of local disorder~\cite{song_2014}, relevant when considering imperfections in experimental setups~\cite{tao_2020,rancic_2022,dvir_2023}, effectively modifies $\nu_{LR}$ and may be utilized to stabilize the system. To exemplify our result, we investigate the nonequilibrium behavior of a one-dimensional system initialized in a topological phase~\cite{kitaev_2001}, where both $H^{(i)}$ and $H^{(f)}$ belong to the BDI symmetry class, identified by the presence of PHS, TRS, and ChS~\cite{ryu_2010}. Since the TRS and ChS are dynamically broken~\cite{mcginley_2018}, the symmetry class postquench relaxes from a BDI to a D classification. Moreover, we show that the nonequilibrium setting introduces further modifications to the system and establish a breakdown in SPT order when the entanglement entropy (EE) crosses over from area- to volume-law behavior~\cite{vidal_2003,calabrese_2004,eisert_2010}.

We expect this work to be relevant to ongoing cold atom experiments~\cite{porras_2004,jiang_2011,laflamme_2014}, that are designed with a high degree of controllability and are well isolated from external sources of noise. By focusing on a toy model under specific quench protocols, we identify nontrivial behaviors that may manifest in more complex setups. Further, our exemplification of disorder-preserved topology may inspire future experiments, where long-lived Majorana edge modes are an ongoing research goal.

The outline of our paper is as follows: in \cref{sec:model}, we introduce the free-fermion system used herein, together with an analysis of its topological classification out of equilibrium. In \cref{sec:methods} we present the tools employed in this work to probe topological phases out of equilibrium. This includes a short review of the correlation matrix used throughout this work and an introduction to the pseudospin formalism required for defining a topological invariant in systems without translational invariance. In \cref{sec:probes} we expose the dynamical behavior of the initially localized edge modes, as a motivation for the remainder of the study and analyze in detail the topological properties of this system. We define several timescales exposing the nontrivial dynamics of the system and use the Loschmidt echo as physical probe of these behaviors. This section concludes with an analysis of the entanglement spectrum and the entanglement entropy, conclusively identifying when the SPT phase becomes ill-defined. Finally, in \cref{sec:discussion}, we summarize our results and outline a number of open questions that follow directly from this work.

%%%%%%%%%%%%%%%%%%%%%%%%%%%%%%%%%%%%%%%%%%%%%%%%%%%%%%%%%%%%%%%%%%%%%%%%%%%%%%%%%%%
%%%%%%%%%%%%%%%%%%%%%%%%%%%%%%%%%%%%%% MODEL %%%%%%%%%%%%%%%%%%%%%%%%%%%%%%%%%%%%%%
%%%%%%%%%%%%%%%%%%%%%%%%%%%%%%%%%%%%%%%%%%%%%%%%%%%%%%%%%%%%%%%%%%%%%%%%%%%%%%%%%%%

\section{Model\label{sec:model}}

\subsection{Hamiltonian}

We study a model of spinless fermions on a quantum wire consisting of $N$ sites with open (OBC) and periodic (PBC) boundary conditions, described by the Kitaev chain Hamiltonian~\cite{kitaev_2001},
\begin{align}
    \label{eq:fermion_ham}
    H^{(\alpha)}=&\sum_j
    \left[-J^{(\alpha)}c_j^\dagger c_{j+1} + \Delta^{(\alpha)} c_j c_{j+1} + \mathrm{h.c.}\right]
    \nonumber\\
    &- \mu_j^{(\alpha)}\left[c_j^\dagger c_j-\tfrac{1}{2}\right],
\end{align}
with $c_j^\dagger$ and $c_j$ being the fermionic creation and annihilation operators, respectively, at site $j$. These operators obey the usual anticommutation relations, $\{c_i,~c_j\}=\{c_i^\dagger,~c_j^\dagger\}=0$ and $\{c_i,~c_j^\dagger\}=\delta_{ij}$, with $J^{(\alpha)}\geq0$ the nearest-neighbor hopping amplitude, $\Delta^{(\alpha)}=e^{-i\theta}|\Delta^{(\alpha)}|$ the $p-$wave superconducting pairing amplitude, $\mu^{(\alpha)}_j$ the local chemical potential, and $\alpha\in(i, f)$ identifying the initial and postquench parameters. Throughout, we employ energy and time units of $J^{(i)}$ and $\hbar/J^{(i)}$, respectively (henceforth setting $\hbar=1$). In the case of OBC we identify $c_{N+1}=0$, whereas for PBC we identify $c_{N+1}=c_1$.

To probe the dynamics of this system in the presence of disorder, we investigate the time evolution of single-particle states up to half-filling, initialized in the nontrivial topological phase $(J^{(i)},~\Delta^{(i)},~\mu^{(i)}_j)=(1,~0.9,~0.2)$. In $H^{(f)}$, the chemical potential is modified by a local disorder $\mu^{(f)}_j=\mu^{(f)}+w_j$, where $w_j\in [-W, W]$ is a uniformly distributed random number with $W\geq0$~\cite{anderson_1958}. We identify the impact of disorder by averaging over $100$ distinct realizations.

\subsection{Topological classification out of equilibrium}

The Kitaev chain described by Eq.~\eqref{eq:fermion_ham} at equilibrium is classified by a bulk topological invariant $\nu_\mathrm{BDI}\in\mathbb{Z}$, where $|\nu_\mathrm{BDI}|$ identifies the number of exponentially localized Majorana zero-modes (MZMs) at either end of an open chain~\cite{chiu_2016}. The nontrivial phases corresponding to $|\mu|<2J^{(i)}$ have invariants $\nu_\mathrm{BDI}=\pm 1$ which, due to bulk-boundary correspondence, identifies a pair of chiral MZMs in each phase.

In this work, we investigate quenches likewise defined by the Hamiltonian of Eq.~\eqref{eq:fermion_ham}. As such, $H^{(f)}$ shares the same discrete symmetries as the prequench Hamiltonian and, thus, also lies in the BDI symmetry class and is classified by $\nu_\mathrm{BDI}$. However, the postquench system is classified not by the symmetries of $H^{(f)}$, but rather by the symmetries present in the time-evolved fictitious Hamiltonian, $H_\mathrm{fic}(t)=U(t)H^{(i)}U^\dagger(t)$, where $U(t)=e^{-iH^{(f)}t}$. Due to the unitarity of the time-evolution operator, Hamiltonians $H^{(i)}$ and $H_\mathrm{fic}(t)$ share the same gapped energy spectrum; however, ChS and TRS are not preserved for $t>0$. This symmetry reduction modifies the classification of the postquench system to the D class, which is described by a $\nu_D\in\mathbb{Z}_2$ topological invariant~\cite{mcginley_2019}. The system therefore seemingly loses information regarding the chirality of the system and retains only binary information regarding the phase: topological or trivial, which may be mediated in this setup by a dynamical change in the symmetries.

%%%%%%%%%%%%%%%%%%%%%%%%%%%%%%%%%%%%%%%%%%%%%%%%%%%%%%%%%%%%%%%%%%%%%%%%%%%%%%%%%%%
%%%%%%%%%%%%%%%%%%%%%%%%%%%%%%%%%%%%% SETTING %%%%%%%%%%%%%%%%%%%%%%%%%%%%%%%%%%%%%
%%%%%%%%%%%%%%%%%%%%%%%%%%%%%%%%%%%%%%%%%%%%%%%%%%%%%%%%%%%%%%%%%%%%%%%%%%%%%%%%%%%

\section{Methods\label{sec:methods}}

It has been established that SPT states in their topological regime support degenerate zero energy modes bound to defects or at the boundaries of the lattice. In a 1D chain of spinless fermions these modes manifest as fractionalized Majorana fermions, which may be expressed in terms of the complex fermionic creation and annihilation operators as $c^\dagger_j=\frac{1}{2}(\gamma_{2j}+i\gamma_{2j+1})e^{i\theta/2}$ and $c_j=\frac{1}{2}(\gamma_{2j}-i\gamma_{2j+1})e^{-i\theta/2}$, respectively, and satisfy the relations: $\{\gamma_j, \gamma_k\}=2\delta_{jk}$, $\gamma_j=\gamma_j^\dagger$ and ${(\gamma_j)}^2=1$. Using these two complementary representations, we shall now introduce the key tools utilized throughout this work.

\subsection{Correlation matrix}

Throughout our numerical studies, we have made extensive use of the correlation matrix, $C$, which contains information about all quantum correlations present in a noninteracting system. For an instantaneous many-body state, $\ket{\Psi(t)}$, this can be written in the Majorana representation as $C_{jk}(t)=\braket{\Psi(t)|\gamma_j\gamma_k|\Psi(t)}$, a $2N\times2N$ matrix where $N$ is the number of sites in the system and $j,k$ run over the $2N$ distinct Majorana operators. To express it in a more useful form, one needs to solve the eigenproblem of $\mathcal{H}$, the single-particle Hamiltonian in the Majorana representation: $H=\sum_{ij}\mathcal{H}_{ij}\gamma_i \gamma_{j}$. With the corresponding instantaneous eigenfunctions, $\psi_m(t)$, the correlation matrix is given by $C_{jk}(t) =\delta_{jk} +i\Gamma_{jk}(t) = \delta_{jk} +2i\text{Im}\sum_m \ket{\psi_m(t)}\bra{\psi_m(t)}_{jk}$, where $m$ extends over the occupied single particle states. In the particular case of half-filling, the expression can be simplified to $C_{jk}(t)=2\sum_m \ket{\psi_m(t)}\bra{\psi_m(t)}_{jk}$. For the interested reader, in \cref{appendix:correlation_matrix} we elaborate on the details of the derivation and discuss its relation to the fermionic representation.

\subsection{Dynamics of the pseudospin}

To introduce the tools used to analyze the pseudospin dynamics, let us first consider a system with PBC. In this case, the momentum space representation serves the practical purpose of indexing the single-particle eigenstates of the system. By introducing the Fourier transformed fermionic operators
\begin{align}
    \wt{c}_k = \frac{1}{\sqrt{N}}\sum_je^{ikj}c_j, & & \wt{c}^\dagger_k = \frac{1}{\sqrt{N}}\sum_je^{-ikxj}c_j^\dagger,
\end{align}
the Hamiltonian can be written as a sum of $2\times2$ matrices: $H^{(i/f)}=\sum_k\Phi^\dagger_k\wt{\mathcal{H}}^{(i/f)}(k)\Phi_k$, i.e.~it is partially diagonalized by introducing the momentum space spinor field $\Phi_k=( \wt{c}_k,\,\wt{c}^\dagger_{-k})$. Then, for any $k$ in the Brillouin zone,
\begin{align}
    \wt{\mathcal{H}}(k)=&d\left(\begin{array}{cc} n_z & n_x-in_y \\ n_x+in_y & -n_z\end{array}\right)\nn\\
    =&d\left( n_x,\, n_y ,\, n_z\right)\cdot\bm{\sigma}\equiv \ve{d}(k)\cdot\bm{\sigma},\label{eq:kspaceH}
\end{align}
where length $d=|\ve{d}|$ is the absolute value of the energy, $\ve{n}=\frac{\ve{d}}{d}$ is the \textit{pseudospin}, and $\ve{\sigma}=(\sigma_x, \sigma_y, \sigma_z)^T$ is a vector containing the Pauli-matrices. There are two eigenvectors with energies $\pm d$, corresponding to the \textit{Bloch bands} of particles and holes, respectively: $\ve{d}\cdot\bm{\sigma}\left|k,\pm\right> =\pm d \left|k,\pm\right>$:
\begin{align}
    \ket{k,+}&=\frac{1}{\sqrt{\tilde{N}}}\left(\begin{array}{c}a \\ b\end{array}\right),\quad
    \ket{k,-}&=\frac{1}{\sqrt{\tilde{N}}}\left(\begin{array}{c} -b^* \\ a\end{array}\right),
\end{align}
where $a=1+n_z$, $b=n_x+in_y$, and $\tilde{N}=2(1+n_z)$. These are orthogonal, so one can construct the projectors to the upper and lower Bloch band, respectively, as: $\ket{k,\pm}\bra{k,\pm}=\frac{1}{2}\left(\mathbbm{1}_2\pm\ve{n}(k)\cdot\bm{\sigma}\right)$.

To track the time-evolution of the pseudospin, we define the \textit{pseudospin operator} in the Heisenberg picture as $\ve{s}(k,t) =e^{it\wt{\mathcal{H}}^{(f)}(k)} \bm{\sigma} e^{-it\wt{\mathcal{H}}^{(f)}(k)}$. Its expectation value with respect to a negative energy eigenstate of $\wt{\mathcal{H}}^{(i)}(k)$ with a given momentum $k$ gives $\left<k,-\right|\ve{s}(k,t)\left|k,-\right> = -\ve{n}(k,t)$. It is possible to derive a closed expression for the pseudospin by solving the differential equation $\partial_t\ve{n}(k,t)=2\ve{d}^{(f)}(k)\times\ve{n}(k,t)$ with the initial value $\ve{n}(k,t=0)=\ve{n}^{(i)}(k)$, which results in:
\begin{align}
    \ve{n}(k,t) =&
    \cos 2\vartheta~\ve{n}^{(i)}+\sin 2\vartheta~\ve{n}^{(f)}\times\ve{n}^{(i)}\nn\\
    +&2\sin^2\vartheta~\ve{n}^{(i)}\cdot\ve{n}^{(f)} \,\ve{n}^{(f)},\label{eq:psSpinVec}
\end{align}
where $\vartheta=|\ve{d}^{(f)}(k)|t$. The pseudospin vectors corresponding to the pre- and postquench Hamiltonians are defined as $\ve{n}^{(i/f)}(k):=\frac{\ve{d}^{(i/f)}(k)}{|\ve{d}^{(i/f)}(k)|}$, cf. \cref{eq:kspaceH}. For illustration, let us assume the hopping, chemical potential and pairing amplitude in the initial and final systems to be $(J,\,0,\,\Delta)$ and $(J,\,\mu,\,0)$, respectively. The corresponding time-dependent pseudospin texture then reads as follows:
\begin{align}
    \ve{n}(k,t) =& \frac{-1}{\sqrt{\Xi}}\left(
    \begin{array}{c}
        \Delta\sin k\sin\left(2t(J\cos k +\mu)\right) \\
        \Delta\sin k\cos\left(2t(J\cos k +\mu)\right) \\
        J\cos k
    \end{array}
    \right),\label{eq:psSpinVecEx}
\end{align}
where $\Xi=J^2\cos^2 k +\Delta^2\sin^2 k$. One can immediately see that $\ve{n}(k,t>0)$ is no longer restricted to the $yz-$plane, as is the case for both $\ve{n}^{(i)}$ and $\ve{n}^{(f)}$. Later we will show how the pseudospin dynamics are reflected in the topological number for the initial BDI system, $\nu_\mathrm{BDI}$.

Before examining the relationship between pseudospin and the previously defined correlation matrix, let us briefly discuss its role in a nonperiodic system. The simplifications coming with the usual momentum space are lost in the nonperiodic case, i.e.~one needs two momenta to parametrize a quadratic Hamiltonian since the energy does not only depend on the position difference of the spinors but their actual position as well. Analogous to the pseudospin operator, one can then define $\bm{\mathfrak{s}}(k,k'):=\Phi^\dagger_k\bm{\sigma}\Phi_{k'}$ by the Fourier transform of the nonlocal coordinate space operator, $\bm{\mathfrak{s}}(x,x'):=\Phi^\dagger(x)\bm{\sigma}\Phi(x')$. To describe correlations between coordinates $x$ and $x'$ one can use an ``average'' position, $X=\frac{x+x'}{2}$, and $y=x-x'$ the deviation from it. Taking the Fourier transform with respect to $y$ gives the ``usual'' momentum variable in periodic systems, but now the resulting quantity retains a dependence on the coordinate $X$. We note here, that since $\avr{\bm{\mathfrak{s}}(x,x')}$ is not assumed to be periodic, there is only one-to-one correspondence between the Fourier-transformed and real-space quantities in the infinite-system-size (or continuum) limit. Working with the Fourier transform of the generalized pseudospin:
\begin{align}
    &\sum\limits_{y}e^{iky}\sum\limits_X \avr{\bm{\mathfrak{s}}(X+y/2,X-y/2)}\nn\\
    &=\frac{1}{N}\sum_{X,y}\sum_{q,q'} e^{-iX(q-q')+iy\left(k-\frac{q+q'}{2}\right)}\avr{\Phi^\dagger_q\bm{\sigma}\Phi_{q'}}\nn\\
    &=\avr{\Phi^\dagger_k\bm{\sigma}\Phi_k} \equiv -\ve{n}(k),
\end{align}
where we assumed that in the initial state of the system, all negative energy states are occupied. One realizes that $\ve{n}(k)$ is the spatial average of the \textit{local pseudospin} $\ve{n}(X,k)=-\sum_y e^{iky}\avr{\Phi^\dagger(X+\frac{y}{2})\bm{\sigma}\Phi(X-\frac{y}{2})} $. Consequently, one can introduce the \textit{effective Bloch bands} as the eigenvectors of the average pseudospin matrix: $\ve{n}(k)\cdot\bm{\sigma} =:\ket{k,+}\bra{k,+}-\ket{k,-}\bra{k,-}$. For a finite-size system this relationship is not exact, however $\ve{n}$ is still a meaningful quantity which approximates the spatially averaged pseudospin of the system. The pseudospin dynamics in the inhomogeneous case, including the equation of motion for the spin texture, can be found in Appendix~\ref{sup:psSpinDynamics}.

\subsection{Pseudospin from the correlation matrix}
As we have previously seen, the expectation value $\avr{\Phi^\dagger_k(t)\bm{\sigma}\Phi_k(t)}=-\ve{n}(k,t)$ gives the pseudospin texture. Here we express $\ve{n}$ with the help of the single-particle correlation matrix. After rearranging the trace, one has
\begin{align}
\ve{n}^{(M)}(k,t) =&-\frac{1}{2N}\sum_{jj'}e^{ik(j-j')}\text{tr}\left\{\bm{\sigma}\widehat{C}_{jj'}(t)\right\},
\end{align}
where $\widehat{C}_{jj'}(t)$ is the correlation matrix in the Majorana representation, which for a given $jj'$ also contains the expectation values $\avr{\gamma_j \gamma_{j'}}$, $\avr{\gamma_{j} \gamma_{j'+1}}$, $\avr{\gamma_{j+1} \gamma_{j'}}$ and $\avr{\gamma_{j+1} \gamma_{j'+1}}$ arranged in a two-by-two matrix and $j$, $j'$ running from 0 to $N$. This is equivalent to the previous expression we have from the fermionic representation up to a permutation of the vector components: $\ve{n}^{(M)}=(n_y,n_z,n_x)$. Using the symmetry properties of the correlation matrix, detailed in \cref{sup:psPsinCorrMx}, we finally arrive at the following expressions:
\begin{align}
 n_x%(k,t)
 =&\sum_{jj'}\sin(k(j-j'))\frac{\Gamma_{2j+1,2j'+1}(t) -\Gamma_{2j,2j'}(t)}{2N}, \label{eq:psSpinX}\\
 n_y%(k,t)
 =& \sum_{jj'}\sin(k(j'-j))\frac{\Gamma_{2j+1,2j'}(t) +\Gamma_{2j,2j'+1}(t)}{2N}, \label{eq:psSpinY}\\
 n_z%(k,t)
 =& \sum_{jj'}\cos(k(j-j'))\frac{\Gamma_{2j+1,2j'}(t) -\Gamma_{2j,2j'+1}(t)}{2N}.\label{eq:psSpinZ}
 \end{align}
The structure of the formulas above is determined by PHS through the form of the spinors. Using the antisymmetry of $\Gamma_{jj'}$ in its indices, it is straightforward to show that $n_x(-k,t)=-n_x(k,t)$, $n_y(-k,t)=-n_y(k,t)$ and $n_z(-k,t)=n_z(k,t)$. We shall use the above direct connection between the trace-less part of the Majorana correlation matrix $\Gamma_{jj'}(t)$ and the pseudospin texture given by $n_{x,y,z}(t)$ to characterize the postquench bulk dynamics of the system.

\subsection{Relation to the topological invariant}\label{sec:topoInv}
For 1D noninteracting fermionic systems in the BDI symmetry class the $\mathbb{Z}$-valued topological invariant can be expressed in terms of the single-particle Green's function, $G(\omega,k)=[i\omega-\wt{\mathcal{H}}(k)]^{-1}$, which is related to the inverse of the Hamiltonian in the reciprocal space for periodic systems. Following Refs.~\cite{gurarie_2011,manmana_2012} we write the topological invariant as
\begin{align}
    \nu_\mathrm{BDI} =\frac{1}{4\pi i}\intlim{k}{\mathrm{BZ}}{}\text{tr}\left\{ \Sigma g^{-1}(k)\partial_k g(k) \right\},
\end{align}
where $\Sigma$ represents the chiral symmetry on the spinors in momentum space, $\Sigma=\sigma_x$, $g(k)=G(\omega =0,k)$. We define the topological invariant for the disordered case analogous to the clean system: considering the spatially-averaged Green's function, which is directly related to the average pseudospin~\cite{gong_2018,chang_2018,yang_2018,hsu_2021}. Employing straightforward algebraic steps, we arrive at
\begin{align}
    \nu_\mathrm{BDI}=&\frac{1}{4\pi i}\intlim{k}{\mathrm{BZ}}{}\text{tr}\left\{\sigma_x \ve{n}(k)\cdot\bm{\sigma} \,\partial_k\ve{n}(k)\cdot\bm{\sigma} \right\}\nn\\
    =&\frac{1}{2\pi}\intlim{k}{\mathrm{BZ}}{}\frac{n_y\partial_kn_z-n_z\partial_kn_y}{n_y^2+n_z^2} \equiv \text{WN}_x\nn\\
    =&\frac{1}{2\pi}\intlim{k}{\mathrm{BZ}}{}\frac{n_y\partial_kn_z-n_z\partial_kn_y}{n_x+1}.\label{equ:nuBDI}
\end{align}
The topological invariant equals to the winding number of the curve $k\mapsto\ve{n}(k)$ around the $x-$axis, $\text{WN}_x$. In the last step we used the direct consequence of PHS: $n_{x,y}$ are odd, while $n_z$ is even function of $k$ --- as we have indicated in the previous subsection.

One can repeat the above argument in the Majorana representation where $\Sigma=\sigma_z$, leading to the equality of $\nu_\mathrm{BDI}$ and the $\text{WN}_z$~\cite{mcginley_2019} of the Majorana pseudospin texture $\ve{n}^{(M)}(k)$. This also allows us to realize the topological invariant as the Zak-phase or 1D Chern-Simons invariant~\cite{mcginley_2018} of the Majorana Bloch band wavefunction $\ket{\phi_k}$:
\begin{align}
    \nu_\mathrm{BDI} =& \frac{1}{2\pi}\intlim{k}{\mathrm{BZ}}{}\frac{n^{(M)}_x\partial_k n^{(M)}_y-n^{(M)}_y\partial_k n^{(M)}_x}{n^{(M)}_z+1}\nn\\
    =&\frac{i}{\pi}\intlim{k}{\mathrm{BZ}}{}\braket{\phi_k|\partial_k\phi_k} = \frac{\alpha}{\pi} =2\mathrm{CS}_1,
\end{align}
with the Zak-phase $\alpha$. For $t>0$, we use the instantaneous value of $\mathrm{CS}_1(t)$ to characterize the topological state of the system ~\cite{zak_1989,ryu_2010,budich_2013,mcginley_2018,rahul_2019} by substituting the effective Bloch band with its time-evolved value $\ket{\phi_k(t)}$.

Although the postquench system can no longer be classified as BDI, and hence $\nu_\mathrm{BDI}$ cannot accurately track its topological state, the time-dependence of $\mathrm{CS}_1(t)$ still carries information about the dynamics of the pseudospin texture. For illustration, let us consider its value for the quench already mentioned in the previous subsection, resulting in \cref{eq:psSpinVecEx}. We are interested in the case when the postquench chemical potential $\mu$ is much larger than any other energy scales: this can also give us a hint regarding the dynamics in the large disorder limit, since in that case the on-site potential dominates over the hopping at every point of the chain. After substituting into the expression, \cref{equ:nuBDI}, we have:
\begin{align}
    \mathrm{CS}_1(t) \underset{J=\Delta}{\overset{\mu\gg J}{\longrightarrow}}& -\frac{1}{2\pi}\intlim{k}{0}{\pi}\frac{\cos(2t\mu) +2tJ\sin^2k\cos k\sin(2t\mu)}{1-\sin^2(2t\mu)\sin^2 k}\nn\\
     =&-\frac{1}{2}\text{sgn}\left(\cos(2t\mu)\right),
\end{align}
where we have chosen to set $J=\Delta$ for brevity, though the derivation can also be performed in the general case. The value of $\mathrm{CS}_1$ oscillates between $\pm\frac{1}{2}$, with frequency $2\mu$. This simple exercise also explains, qualitatively, why the onset time $\overline{t}_\text{osc}$ of the $\mathrm{CS}_1$ oscillation is inversely proportional to the disorder strength $W$ --- at least in the large disorder case, when we can assume that the dominant dynamical timescale is set by $W\gg J,\Delta$.

Thus, analyzing the topological phase of the system is equivalent to the classification of curves on the surface of the unit sphere defined by the map $k\mapsto \ve{n}(k,t)$. The pseudospin dynamics offer an intuitive means to understand the difference between the BDI and D class systems. At $t=0$, the initial Hamiltonian restricts $\ve{n}(k,0)$ to lie in the $yz-$plane, where nontrivial phases correspond to curves which wind around the origin clockwise ($\nu_\mathrm{BDI}=1$) or anticlockwise ($\nu_\mathrm{BDI}=-1$). The $k\mapsto \ve{n}(k)$ curves corresponding to a nontrivial topological phase intersect both poles $n_z=\pm 1$ of the unit sphere. Since it is not possible to deform two such nontrivial curves with differing winding number $\text{WN}_x$ into each other while constrained in $yz$, the possible topological phases are classified by $\mathbb{Z}$. In the case of D class systems, for $t>0$ the whole surface of the unit sphere is available for the pseudospin texture $\ve{n}(k,t)$. Two curves belonging to distinct topological phases in the BDI class, for example $\nu_\mathrm{BDI}=\pm 1$, can be members of a family of curves connected by unitary time-evolution. Above we have seen an explicit example of this: identifying oscillations in $\mathrm{CS}_1$ (i.e.~distinct BDI-phases) yet no change in the D class topological index $\nu_D=2(\mathrm{CS}_1\text{ mod }1)$. In the postquench D system, PHS still constrains $n_x$ and $n_y$ to be odd functions of $k$. As a consequence, curves belonging to the topologically nontrivial phase intersect the poles of the unit sphere at $n_z=\pm 1$, while in the trivial phase they do not. Therefore, the corresponding topological number can equivalently be given as $\nu_\mathrm{D}(t)=\frac{1}{2}\left[\text{sign}(n_z(k=\pi,t))-\text{sign}(n_z(k=0,t))\right]=2\left(\mathrm{CS}_1~\mathrm{mod}~1\right)$. This number is pinned to its initial value for $t>0$ in the disorder-free limit. Using the formula in \cref{eq:psSpinVec}, and assuming $\ve{d}^{(i/f)}(k) =\left(0,\,\Delta^{(i/f)}\sin k,\,-J^{(i/f)}/2\cos k -\mu^{(i/f)}\right)$, it is straightforward to show that
\begin{align}
    \nu_D =\frac{1}{2}\left(\text{sign}(2J^{(i)}-\mu^{(i)}) + \text{sign}(2J^{(i)}+\mu^{(i)})\right)\nn
\end{align}
indeed depends on the initial state only, classifying the system as nontrivial for $2|J^{(i)}|>|\mu^{(i)}|$ and trivial for $2|J^{(i)}|<|\mu^{(i)}|$. One might reasonably expect this behavior to persist even for the disordered system. This intuition turns out to be correct for certain quenches, when there are no extended regions in the bulk which could host topologically different phases from others. However, it can still happen that the topological number changes in time for a given realization of the disordered chain. In \cref{sup:psPsinCorrMx} we detail the dynamics of the pseudospin texture in the presence of disorder, which turns out to be dependent not only on the initial configuration, but the history of the system as well. Nevertheless, our numerical investigation suggests that the disorder-averaged topological number $\overline{\nu_D}$ does not change its value during the time-evolution of the system, despite the fact that its dynamics on the single-realization level is more complicated than in the case of the clean system.

%%%%%%%%%%%%%%%%%%%%%%%%%%%%%%%%%%%%%%%%%%%%%%%%%%%%%%%%%%%%%%%%%%%%%%%%%%%%%%%%%%%
%%%%%%%%%%%%%%%%%%%%%%%%%%%%%%% TOPOLOGICAL ANALYSIS %%%%%%%%%%%%%%%%%%%%%%%%%%%%%%
%%%%%%%%%%%%%%%%%%%%%%%%%%%%%%%%%%%%%%%%%%%%%%%%%%%%%%%%%%%%%%%%%%%%%%%%%%%%%%%%%%%

\section{Probing Quench Dynamics\label{sec:probes}}

\subsection{Localization}

\begin{figure}
    \centering
    \includegraphics[width=\linewidth]{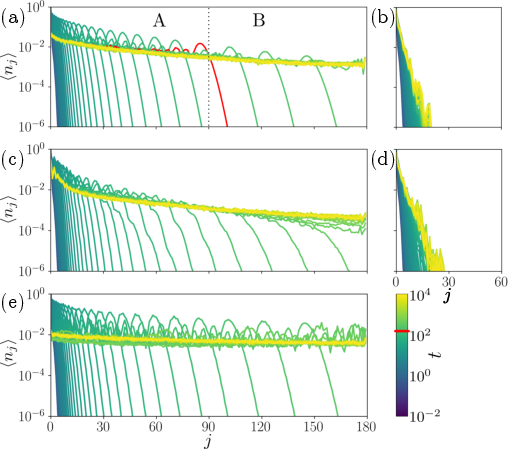}
    \caption{%
        \label{fig:localization}Time evolution of the realization-averaged population profiles $\braket{n_{j}}$, plotted as a function of position $j$, for the MZM initially localized at the left edge of an $N=180$ site chain with OBC. The chain is initialized in the topological regime at $t=0$ with $(J^{(i)},~\Delta^{(i)},~\mu_j^{(i)})=(1,~0.9,~0.2)$. The quench is then defined as $(J^{(f)},~\Delta^{(f)},~\mu_j^{(f)})=(J^{(i)},~\Delta^{(f)},~\mu_0+w_j)$ with parameters: (a,~b) $\Delta^{(f)}=0$ and $\mu_0=0.2$ with $W=0.5$ and $W=5.0$, respectively, (c,~d) $\Delta^{(f)}=0.2$ and $\mu_0=0.2$ with $W=0.5$ and $W=5.0$, respectively, and (e) $\Delta^{(f)}=0.2$, $\mu_0=2.5$ and $W=0.2$. The red curve in panel (a) indicates the profile of $\braket{n_{j}}$ as its traveling peak crosses the entanglement cut at $t\sim180$.
    }
\end{figure}

To address what is happening to the MZMs as they evolve in time, we first calculate the realization-averaged contribution to the local fermion number, $\braket{n_{j}(t)}$, for the initially left-localized single-particle MZM~\footnote{while for an individual realization the left- and right-localized MZMs will, in general, have different profiles, on average we find that they qualitatively exhibit the same dynamics. As such, we choose to discuss solely the left-localized modes for brevity.}. Figure~\ref{fig:localization}~(a) shows that, for a critical quench with $\Delta^{(f)}=0$ and small $W$, the initially localized modes rapidly extend away from the boundary. Further, the leading edge of the wavefront features a prominent peak which travels along the length of the chain before being reflected at the opposite end. This extension of the initially localized mode into the bulk in the presence of small disorder is not unique to critical quenches. Indeed, quenches to the topological regime (defined by the presence of finite $\Delta^{(f)}$ and small $W$) and to the nontopological regime (defined by the presence of finite $\Delta^{(f)}$, small $W$ and $\mu^{(f)}>2J+W$) both exhibit similar profiles after a long time has elapsed, as shown in Figs.~\ref{fig:localization}~(c) and \ref{fig:localization}~(e), respectively. Additionally, it is interesting to note the suppresion of the leading peak in the wavefront when $\Delta^{(f)}$ is comparable to the $\mu^{(f)}_j$ [Fig.~\ref{fig:localization}~(c)], and that large $\mu^{(f)}$ results in the population becoming more evenly distributed at long time when compared to the other low-$W$ quenches since disorder then becomes the smallest energy scale in the system. By contrast, edge mode localization is preserved for large disorder irrespective of the value of $\Delta^{(f)}$ [see Figs.~\ref{fig:localization}~(b) and \ref{fig:localization}~(d)]. Across all regimes, $\braket{n_{j}(t)}$ reaches a steady state after a finite time, with all profiles saturating for times exceeding $t\sim10^3$.

\subsection{Topological Analysis\label{sec:topo}}

\begin{figure}[t!]
    \centering
    \includegraphics[width=\linewidth]{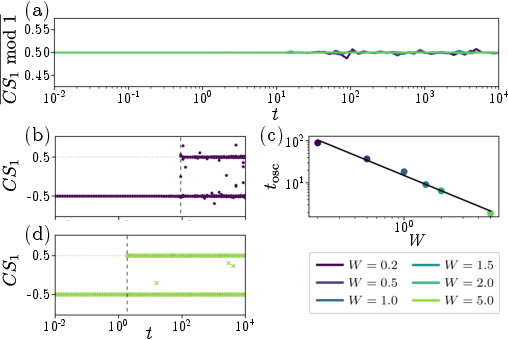}
    \caption{%
        \label{fig:chern_simons}(a) Time evolution of $\overline{\mathrm{CS}_1~\mathrm{mod}~1}$ for an $N=180$ site chain with PBC for a range of $W$. All other quench parameters are the same as in Figs.~\ref{fig:localization}~(a) and \ref{fig:localization}~(b). (b, d) All individual realizations of $\mathrm{CS}_1$ for $W=0.2$ and $W=5.0$, respectively. To resolve $\mathrm{CS}_1$ with minimal noise, $5\times10^4$ $k$-points are used for $W=5$ while $2\times10^5$ are used for $W=0.2$. (c) The time at which oscillations in $\mathrm{CS}_1$ begin, $t_\mathrm{osc}$, [identified by dashed vertical lines in panels (b, d)] varying with $W$. The line in panel (c) indicates a fitting to  $t_\mathrm{osc}=AW^{-b}$ with $A=14.9$ and $b=1.21$.
    }
\end{figure}

To characterize the bulk topological phase of the system with PBC, we calculate the evolution of the one-dimensional Chern-Simons number~\cite{zak_1989,ryu_2010,budich_2013,mcginley_2018,rahul_2019}, introduced previously in Eq.~\ref{equ:nuBDI},
\begin{align}
    \mathrm{CS}_1(t)=\frac{i}{2\pi}\int_\mathrm{BZ}dk\braket{\phi_k(t)|\partial_k\phi_k(t)},
    \label{eq:cs1}
\end{align}
which classifies systems in one spatial dimension~\cite{ryu_2010}.
Here, $\ket{\phi_k}$ are the instantaneous effective Bloch wavefunctions of the occupied bands in the Majorana representation defined in \cref{sec:methods}. In the BDI class, $\mathrm{CS}_1$ is quantized to nonzero half-integer values in the nontrivial phase and is related to the topological invariant via $\nu_\mathrm{BDI}=2\mathrm{CS}_1$. For the Kitaev chain, this is equivalent to the winding number~\cite{mcginley_2019}. postquench, the invariant becomes $\nu_\mathrm{D}(t)=2\left(\mathrm{CS}_1~\mathrm{mod}~1\right)$ since gauge transformations of $\ket{\phi_k}$ can modify the value of $\mathrm{CS}_1$ by an integer~\cite{auckly_1994,budich_2013,mcginley_2018}. Although the presence of disorder in $H^{(f)}$ breaks translational invariance, leaving the momentum-space representation ill-defined, we can still analyze the system in terms of the pseudospin~\cite{gong_2018,chang_2018,yang_2018,hsu_2021} $\bm{n}(k,t)=\braket{\Phi^\dagger_k\bm{\sigma}\Phi_k}$, which has been defined previously and may be extracted from the correlation matrix.

Figure~\ref{fig:chern_simons}~(a) depicts $\overline{\mathrm{CS}_1~\mathrm{mod}~1}$, the mean value of Eq.~\eqref{eq:cs1} modulo $1$ across all realizations. Although the constant value of $\overline{\mathrm{CS}_1~\mathrm{mod}~1}$ indicates that the postquench system remains in the nontrivial phase, other observables, such as the underlying pseudospin, still dynamically evolve. Individual values of $\mathrm{CS}_1$, though initially fixed to $-\sfrac{1}{2}$ across all $W$, begin to oscillate between $\pm\sfrac{1}{2}$ [see Figs.~\ref{fig:chern_simons}~(b) and \ref{fig:chern_simons}~(d)] after a characteristic time which is related to the strength of the disorder via an inverse power law, $\bar{t}_\mathrm{osc}\propto W^{-b}$, where $b\sim1$ [see Fig.~\ref{fig:chern_simons}~(c)]. Although the frequency of these oscillations is not well-defined in the presence of finite disorder, they recover a regular period when the disorder is removed [see Fig.~\ref{fig:return_rate}~(d)]. Opposite signs in $\mathrm{CS}_1$ are associated with different phases in the static BDI system, which are topologically indistinct in the D class out of equilibrium as previously mentioned in \cref{sec:topoInv}.

\subsection{Loschmidt Echo}

\begin{figure*}[t!]
    \centering
    \includegraphics[width=\linewidth]{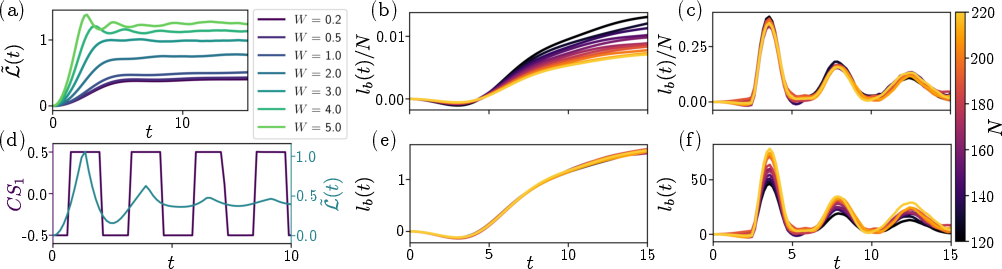}
    \caption{%
        \label{fig:return_rate}(a) Dynamics of the realization-averaged return rate $\tilde{\mathcal{L}}(t)$ for a short period after a quench defined by the same parameters as was employed in Figs.~\ref{fig:localization}~(a) and \ref{fig:localization}~(b) for a range of different $W$. (b, c) The rescaled boundary contribution to the return rate $l_b(t)/N$ using the same quench parameters with $W=0.2$ and $W=5.0$, respectively, for chains with lengths between $N=120$ and $N=220$ in intervals of $10$. (d) Left axis: Oscillations in $\mathrm{CS}_1$ for a quench to the nontopological regime with $\mu_j^{(f)}=5$ and zero-disorder [other values as in Figs.~\ref{fig:localization}~(a) and \ref{fig:localization}~(b)]. Right axis: The corresponding return rate, $\tilde{\mathcal{L}}(t)$. (e, f) The un-scaled boundary contribution to the return rate $l_b(t)$ for $W=0.2$ and $W=5.0$, respectively.
    }
\end{figure*}

To further probe these oscillations and their indication that the system periodically transitions between the previously distinct topological phases we investigate the Loschmidt Echo (LE)~\cite{fisher_1965,fisher_1967,jalabert_2001,vajna_2015,budich_2016,zhang_2020} which quantifies the revival of a quantum state during time evolution, as well as how sensitive the system dynamics are to quantum perturbations~\cite{jalabert_2001}. The LE is defined as $\mathcal{L}(t)=|\braket{\Psi(0)|U(t)|\Psi(0)}|^2$, where $\ket{\Psi(0)}$ is an initial many-body state and $U(t)\ket{\Psi(0)}$ is its time evolved counterpart. In general, for a sufficiently large system, the overlap between the initial and evolved states will become exponentially small for long time, referred to as the orthogonality catastrophe~\cite{anderson_1967}. For this reason it is instead helpful to refer to the return rate (RR), $\tilde{\mathcal{L}}(t)=-\sfrac{1}{N}\log|\mathcal{L}(t)|$, which is analogous to the free energy of the partition function~\cite{sedlmayr_2018,sedlmayr_2019}. In the limit $N\rightarrow\infty$, the LE vanishes at times referred to as Fisher zeros. These, in turn, manifest as sharp peaks in the profile of the RR and are a signature of dynamical phase transitions~\cite{vajna_2015,heyl_2018,pastori_2020,vanhala_2023}.

For a noninteracting system such as the one currently being explored, the LE can alternatively be determined by employing the correlation matrix, $C$, via $\mathcal{L}(t)=|\det(I-C(0)+C(0)U(t))|^2$~\cite{sedlmayr_2018}. Here, $C(0)$ refers to the prequench correlation matrix constructed from all states at $t=0$ up to half-filling, with the time evolution of $\mathcal{L}(t)$ being determined solely by the operator $U(t)=e^{-iH^{(f)}t}$.

The dynamics of the realization-averaged RR exhibit markedly different behavior depending on the scale of the disorder in the system. In Fig.~\ref{fig:return_rate}~(a), we illustrate, for a chain with PBC and the same quench parameters as were employed in Fig.~\ref{fig:chern_simons}, how the presence of large $W$ reduces the overlap with the original state when compared to the profiles for lower $W$ (since high RR corresponds to a lower LE). Additionally, we observe that, as we increase disorder, peaks in the profile of the RR become more prominent and occur at increasingly short times after the quench. Indeed, for large $W$ the time at which the first peak in $\tilde{\mathcal{L}}(t)$ occurs in Fig.~\ref{fig:return_rate}~(a), $t\sim2$, coincides with the time at which oscillations in Fig.~\ref{fig:chern_simons}~(d) appear, further indicating that these oscillations are associated with dynamical phase transitions between the previously distinct topological phases of the prequench BDI class system. while the peaks become increasingly less prominent as $W$ decreases, they are still present at later times, following the qualitative trend illustrated in Fig.~\ref{fig:chern_simons}~(c).

To more clearly establish the connection between the dynamics in $\mathrm{CS}_1$ and peaks in the RR --- and to further support the assertion that they can be ascribed to revivals of the prequench topological phase --- we consider the evolution of a system with zero disorder. For this purpose, we choose to include a significant overall chemical potential, $\mu_j^{(f)}$, such that oscillations in $\mathrm{CS}_1$ are present for short times after the quench and result in sharp features in $\tilde{\mathcal{L}}(t)$. This is depicted in Fig.~\ref{fig:return_rate}~(d), where we observe that peaks in the RR coincide precisely with oscillations in the associated $\mathrm{CS}_1$~\cite{sedlmayr_2015}. Furthermore, we note that peaks in $\tilde{\mathcal{L}}(t)$ exactly coincide with oscillations in the $xz$-winding, which is $\pi/2$ out of phase with both $\mathrm{CS}_1$ and the $xy$-winding, $\text{WN}_z$ (see \cref{sec:topoInv}). This alignment between the return rate and geometric phases has been the subject of previous discussions~\cite{budich_2016}.

Having established a connection between these two bulk properties, $\mathrm{CS}_1$ and $\tilde{\mathcal{L}}(t)$, we shall now probe the influence of the boundary modes by considering the size scaling of a chain with OBC in the two distinct disorder regimes: $W<J^{(f)}$ and $W>J^{(f)}$. The boundary contribution to the RR, $l_b(t)$, may be determined from the full RR via, $\tilde{\mathcal{L}}(t)\approx\tilde{\mathcal{L}}_0(t)+l_b(t)/N$. Here, the bulk contribution $\tilde{\mathcal{L}}_0(t)$ may be taken to be the RR of an equivalent chain with PBC in the thermodynamic limit~\cite{sedlmayr_2018}. For the data presented in the following discussion, we found a PBC chain of length $N=300$ sufficient to isolate features in $l_b(t)$.

In Figs.~\ref{fig:return_rate}~(b) and \ref{fig:return_rate}~(c) we show the rescaled value of the boundary contribution, $l_b(t)/N$, for $W=0.2$ and $W=5.0$, respectively, while in Figs.~\ref{fig:return_rate}~(e) and \ref{fig:return_rate}~(f) we show the un-scaled value $l_b(t)$ for the same disorder windows. Figure~\ref{fig:return_rate}~(c) shows that, much like for the PBC chain, the boundary modes manifest well-defined peaks in the RR under the influence of high disorder. This suggests that they also participate in the dynamical phase transitions of the postquench system, though their oscillation frequency is approximately half what was observed previously in the PBC chain for the same disorder strength. We also note that, while the rescaled values $l_b(t)/N$ collapse onto a single curve for high disorder, for small disorder profiles become increasingly separated with time [Fig.~\ref{fig:return_rate}~(b)]. Rather, we find that in the small disorder regime, the $l_b(t)$ curves collapse onto one another when there is no size scaling. These contrasting results indicate that, while there exists a universal behavior governing systems with large $W$ which scales with $N$ (meaning that the MZMs remain relevant for all system sizes), this is not true for small $W$. This relationship aligns with the framework previously laid out in Fig.~\ref{fig:localization}, where ever-longer chains would support increasingly extended edge modes after a long time that diverge further and further from the initially localized zero mode profile.

%%%%%%%%%%%%%%%%%%%%%%%%%%%%%%%%%%%%%%%%%%%%%%%%%%%%%%%%%%%%%%%%%%%%%%%%%%%%%%%%%%%
%%%%%%%%%%%%%%%%%%%%%%%%%%%%%%%%%%%% CHIRALITY %%%%%%%%%%%%%%%%%%%%%%%%%%%%%%%%%%%%
%%%%%%%%%%%%%%%%%%%%%%%%%%%%%%%%%%%%%%%%%%%%%%%%%%%%%%%%%%%%%%%%%%%%%%%%%%%%%%%%%%%

\subsection{A memory of polarization}

\begin{figure*}[t]
    \centering
    \includegraphics[width=\linewidth]{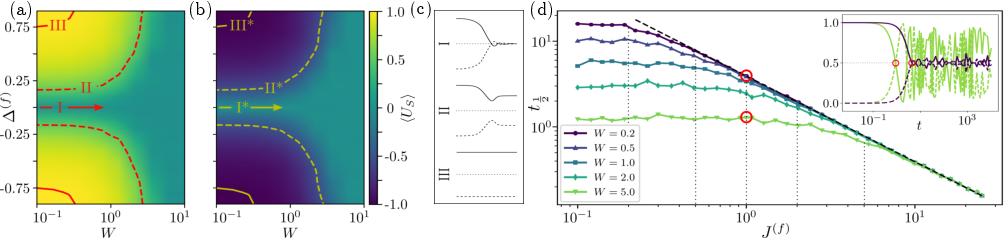}
    \caption{%
        \label{fig:chirality}(a, b) The value to which $\braket{U_S}$ saturates after a long time has elapsed for the MZMs initially localized at the left and right edges, respectively, of an $N=180$ length chain with OBC, averaged over all realizations. $W$ and $\Delta^{(f)}$ are varied, with other parameters the same as in Figs.~\ref{fig:localization}~(a)  and \ref{fig:localization}~(b). (c) Sketch of the realization-averaged Majorana subspecies amplitudes corresponding to contours in (a, b), where $\tilde{n}_{e(o)}$ is represented by the solid (dashed) lines. (d) $t_\frac{1}{2}$ varying with $J^{(f)}$ for $\Delta^{(f)}=0$ and a range of $W$, other parameters are the same as in Figs.~\ref{fig:localization}~(a)  and \ref{fig:localization}~(b). The polynomial regime, proportional to $(J^{(f)})^{-1}$, is illustrated by the black dashed line. Inset: The Majorana distribution $n_{e(o)}$ for even (odd) subspecies, identified by solid (dashed) lines, for a single randomly selected realization with $W=0.5$ and $W=5$ at $J^{(f)}=1$.
    }
\end{figure*}

Together, the observations of the previous sections build a picture of two starkly different dynamical regimes. The first is that of low disorder, in which the degenerate edge modes diverge significantly from their initial profile while there remains a sizable overlap within the bulk modes. In the second, high disorder preserves the localized profile of the MZMs but disrupts the bulk modes, leading to a comparatively lower overlap (LE) with the initial system featuring well-defined Fisher zeros. However, this does little to isolate what is happening at the level of the Majoranas --- since although large disorder keeps the edge mode pinned to the boundary of the chain, it does not preclude oscillation within the same site. Rather, we would expect an enhancement of such oscillations in this instance since the disorder directly couples Majoranas sharing the same spatial location. To this end, we now investigate the evolution of the MZM's chiral polarization.

The prequench system possesses chiral symmetry (ChS), \mbox{$U_S\mathcal{H}_\text{fic}(t=0)U_S^\dagger=-\mathcal{H}_\text{fic}(t=0)$}, with $U_S$ representing the discrete symmetry operator anticommuting with the first-quantized Hamiltonian $\mathcal{H}_\text{fic}$, defined via $H_\mathrm{fic}(t)=\sum_{ij}\mathcal{H}_{\text{fic},ij}(t)\Phi_i^\dagger\Phi_j$. Initially, the MZMs are \textit{polarized} with respect to ChS, in the sense that $\braket{\psi_{\pm}|U_S|\psi_{\pm}}=\pm 1$ for the two zero energy modes $\ket{\psi_{\pm}}$~\footnote{This is a direct consequence of the initial discrete symmetries: $(i)$ TRS makes $\mathcal{H}(t=0)$ real when $\Delta$ is real; $(ii)$ PHS maps any finite energy state into an orthogonal one, except for the Majoranas. The ChS is a composition of TRS and PHS. TRS is represented by complex conjugation because of $(i)$, so the corresponding PHS eigenvalues $\pm 1$ classify the MZMs}. This relation no-longer strictly holds for $t>0$, yet one can still define the MZM polarization $\braket{U_S}_\pm(t)=\braket{\psi_\pm(t)|U_S|\psi_\pm(t)}$ for the time-evolved zero modes.

In the Majorana representation, the ChS operator is given by $U_S=\mathbbm{1}_N\otimes\sigma_z$. Therefore, the MZM polarization may be expressed as $\avr{U_S}_\pm=(\tilde{n}_e-\tilde{n}_o)_\pm$ for a given single-particle state, where $\tilde{n}_{e(o)}(t)$ is the realization-averaged sum over the even (odd) Majorana subspecies, given by $n_e=\sum_n|\psi_{\pm, 2n}|^2$ and $n_o=\sum_n|\psi_{\pm, 2n-1}|^2$. Figures~\ref{fig:chirality}~(a) and \ref{fig:chirality}~(b) show the asymptotic behavior of $\avr{U_S}$ after a long time, identifying a competition between the protection of polarization for nonzero $\Delta^{(f)}$ and the suppression induced by $W$ for both the initially (a) left- and (b) right-localized MZMs. How the profiles of the corresponding Majorana populations evolve may be broadly categorized into the three regimes depicted in Fig.~\ref{fig:chirality}~(c), which illustrate how the realization-averaged even (solid line) and odd (dashed line) Majorana amplitudes vary with time. In regime $I$ polarization equilibrates within a finite time, corresponding to quenches with either $\Delta^{(f)}=0$ or with sufficiently large $W$. Conversely, we find that the edge modes retain perfect polarization when $\Delta^{(f)}=\Delta^{(i)}>0$ and $W\rightarrow0$, illustrated in region $III$. All other quenches fall into region $II$, with the realization-averaged polarization of the MZMs exhibiting oscillations in their trajectories before saturating with partial retention of the initial polarization. Approximate contours corresponding to each of these three regimes are also shown in Fig.~\ref{fig:chirality}~(a), where region $I$ lies along the $\Delta^{(f)}=0$ line and spans much of the right-hand-side at high $W$, region $II$ is illustrated for $\avr{U_S}=0.5$ (though for clarity we reiterate that most of the diagram lies in this regime), and for region $III$ we show the contour corresponding to $\avr{U_S}=0.99$. Starred labels in Fig.~\ref{fig:chirality}~(b) correspond to states which have opposite polarization to those in Fig.~\ref{fig:chirality}~(a).

Figure~\ref{fig:chirality}~(d) shows the value of $t_\frac{1}{2}$, the earliest time satisfying $\tilde{n}_{e(o)}(t_{\frac{1}{2}})=0.5$, for $\Delta^{(f)}=0$ and varying $J^{(f)}$. This quantity is employed here to characterize the polarization decay time. Two distinct regimes may be observed: one where $t_\frac{1}{2}$ is independent of the hopping amplitude, corresponding to $W \geq J^{(f)}$, and the second where $t_\frac{1}{2}\propto(J^{(f)})^{-1}$, corresponding to $W < J^{(f)}$. As was observed in the profiles of $\mathrm{CS}_1$, oscillations again appear in individual realizations of $n_{e(o)}$. We exemplify this in the inset of Fig.~\ref{fig:chirality}~(d) for two values of $W$ at $\Delta^{(f)}=0$, where larger $W$ corresponds to increased amplitude of oscillation. Moreover, the time at which these oscillations begin qualitatively exhibits the same $W$ dependence as $t_\mathrm{osc}$ [Fig.~\ref{fig:chern_simons}~(c)], highlighting another clear divergence from the static case where polarization is pinned to $\pm1$. Next, we probe the system further through an investigation into its ES.

%%%%%%%%%%%%%%%%%%%%%%%%%%%%%%%%%%%%%%%%%%%%%%%%%%%%%%%%%%%%%%%%%%%%%%%%%%%%%%%%%%%
%%%%%%%%%%%%%%%%%%%%%%%%%%%%%% ENTANGLEMENT SPECTRUM %%%%%%%%%%%%%%%%%%%%%%%%%%%%%%
%%%%%%%%%%%%%%%%%%%%%%%%%%%%%%%%%%%%%%%%%%%%%%%%%%%%%%%%%%%%%%%%%%%%%%%%%%%%%%%%%%%

\subsection{Entanglement\label{sec:entanglement}}

\begin{figure*}[th!]
    \centering
    \includegraphics[width=0.75\textwidth]{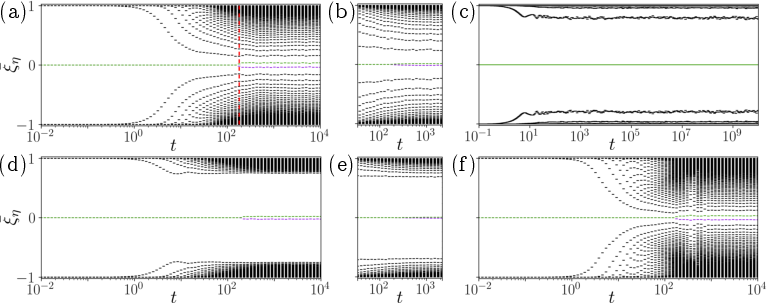}
    \caption{%
        \label{fig:entanglement_spectra}Evolution of the realization-averaged entanglement spectra, $\bar{\xi}_\eta$, for a chain of length $N=180$ with OBC. (a-c) ES for the same quench parameters as used in Figs.~\ref{fig:localization}~(a) and \ref{fig:localization}~(b) with disorder (a) $W=0.5$, (b) $W=1.0$ and (c) $W=5.0$. (d, e) ES for the quench depicted in Figs.~\ref{fig:localization}~(c) and \ref{fig:localization}~(d) with disorder (d) $W=0.5$ and (e) $W=1.0$. (f) ES for the quench parameters used in Fig.~\ref{fig:localization}~(e). Spectra depicted in panels (a, c, d, f) exactly correspond to the population profiles shown in Figs.~\ref{fig:localization}~(a), \ref{fig:localization}~(b), \ref{fig:localization}~(c) and  \ref{fig:localization}~(e), respectively. The red dashed line in panel (a) identifies the time at which a gap opens between the degenerate zero modes, exactly coinciding with the time at which the extended edge mode crosses the entanglement cut in Fig.~\ref{fig:localization}~(a).
    }
\end{figure*}

It is well known that it is possible to characterize a quantum system by its entanglement features. For example: product states, thermal states, localized states and topological phases of matter all have unique signatures diagnosed by the inherent entanglement present in the system. In this work, we consider bipartite entanglement between two spatially separated subsystems $A$ and $B$. First, we assume that the full Hilbert space of a system $\mathcal{H}$ can be expressed as a tensor product of the subsystem Hilbert spaces $\mathcal{H}=\mathcal{H}_A\otimes\mathcal{H}_B$. Therefore, any pure state can be expressed as a tensor product of states in each subsystem's Hilbert space $\ket{\Psi}=\sum_{j,k}\Psi_{j,k}\ket{\Psi_j}_A\!\ket{\Psi_k}_B$. The states $\ket{\Psi_j}_{A(B)}$ form an orthonormal basis in $\mathcal{H}_A(\mathcal{H}_B)$ and the matrix $\Psi_{j,k}$ is, in general, not diagonal. Using a Schmidt decomposition we bring this into diagonal form $\ket{\Psi}=\sum_{j}\alpha_{j}\ket{\alpha_j}_A\!\ket{\alpha_j}_B$, where the Schmidt eigenvalues $\{\alpha_j\}$, satisfying $\alpha_j\geq0$ and $\sum_j\alpha_j^2=1$, carry information about bipartite entanglement in the system. Formally, the entanglement spectrum is related to the Schmidt eigenvalues by $\{-\mathrm{ln}\,\alpha_j\}$, a reparameterization of probabilities to effective energies. A single nonzero $\alpha_j$ indicates a product state with no entanglement, whereas systems with more than one nonzero $\alpha_j$ are entangled. Various parameterizations of $\{\alpha_j\}$ can be employed to expose different underlying entanglement features. In the following, we consider two entanglement measures pertinent to this work: the single-particle ES and the EE.

\subsubsection{Entanglement spectrum}

The ES has found broad use as a tool for categorizing topological systems~\cite{li_2008,legner_2013,patrick_2017,gong_2018,sayyad_2021} where a counting of degeneracies in the low-lying many-body ES is equivalent to a counting of MZMs localized at the boundaries in real space~\cite{li_2008, fidkowski_2010, chandran_2011, gong_2018}. Due to bulk-boundary correspondence, this degeneracy identifies a nontrivial topological invariant and is therefore indicative of a topological phase~\cite{li_2008,pollmann_2010,cho_2017}. Furthermore, it has been suggested that in both the static and dynamic settings, a lifting of the many-body ES degeneracies can be interpreted as an indication of a topological phase transition~\cite{mcginley_2019}. For a noninteracting system the many-body ES can be constructed via sums of single-particle energies. Therefore, identifying degeneracies is equivalent to identifying zeros in the single-particle ES~\cite{peschel_2003,peschel_2009}.

The single-particle ES, $\{\xi_\eta\}$, between subsystems $A$ and $B$ is found by diagonalizing $\Gamma=2C_A-I$, where $C_A=P_ACP_A$ is the reduced correlation matrix projected into the $A$ subsystem by $P_A$~\cite{peschel_2003,peschel_2009,legner_2013}. Figures~\ref{fig:entanglement_spectra}~(a)-\ref{fig:entanglement_spectra}~(c) show the realization-averaged ES for a range of $W$, with $\Delta^{(f)}=0$ and a half-chain bipartition, $N_A=N/2$. For small disorder [Fig.~\ref{fig:entanglement_spectra}~(a)] we initially observe a bulk gap in the ES with gapless modes pinned to zero energy. The time at which the ES zero-energy degeneracy is lifted (vertical red dashed line) corresponds to when the propagating peak in $\braket{n_{j}(t)}$ crosses the boundary between the $A$ and $B$ subdomains [identified by the red curve in Fig.~\ref{fig:localization}~(a)], leading to a finite entanglement between subsystems $A$ and $B$ attributable to the traveling edge mode. At the same time, the bulk gap closes indicating that the edge modes become indistinguishable from the bulk modes. Conversely, in the ES depicted in Fig.~\ref{fig:entanglement_spectra}~(c) for large disorder and evolved over a significantly longer timescale, the zero-energy modes remain pinned to zero and well separated from bulk modes throughout the evolution. This clearly indicates a retention of the prequench localization due to the presence of disorder which freezes the fermion populations and results in the system reaching a steady state that preserves topology [see Fig.~\ref{fig:localization}~(b)]. For intermediate values of disorder comparable to $J^{(f)}$ [Fig.~\ref{fig:entanglement_spectra}~(b)], we find that the splitting of the zero-energy modes is reduced, though still present, and that the bulk modes more slowly approach the mid-gap states. The corresponding population profile exhibits clear evidence of localization through exponential decay from the edge. However, it does so with a far longer localisation length, resulting in finite entanglement at the $A$ and $B$ subdomain boundary.

In Figs~\ref{fig:entanglement_spectra}~(d)-\ref{fig:entanglement_spectra}~(f) we extend this analysis to consider quenches for which $\Delta^{(f)}\neq0$. In Fig.~\ref{fig:entanglement_spectra}~(d), where $\Delta^{(f)}=0.2$ and $W=0.5$ such that the postquench Hamiltonian is gapped and topological, we see that entanglement eigenvalues corresponding to bulk modes remain well separated from those of the edge modes at all times. Nonetheless, the edge modes again induce an ES gap splitting equivalent to that seen in Fig.~\ref{fig:entanglement_spectra}~(a). As was briefly discussed above, this is due to the MZM tails extending across the subsystem partition [see Fig.~\ref{fig:localization}~(c)]. Notably, however, unlike in the discussion above, this now occurs despite both the pre- and postquench Hamiltonians being gapped and in the same topological phase. Under the influence of large disorder (Fig.~\ref{fig:entanglement_spectra}~(e) with $\Delta^{(f)}=0.2$ and $W=5$) the comparatively small $\Delta^{(f)}$ has a negligible effect on the spectrum of $H^{(f)}$ and the topological phase is once again preserved due to the freezing of the fermion populations. Finally, in Fig.~\ref{fig:entanglement_spectra}~(f) we show the ES for a postquench Hamiltonian $H^{(f)}$ which describes a gapped, nontopological phase with $\Delta^{(f)}=0.2$, $W=0.2$, and $\mu^{(f)}=2.5$. In this regime, the initial state of the system is projected onto the eigenstates of the nontopological postquench Hamiltonian. It thus necessarily loses its topology, evidenced by the loss of localization in Fig.~\ref{fig:localization}~(e), with the zero-energy mode degeneracy lifting as the bulk gap simultaneously closes.

\subsubsection{Entanglement entropy}

\begin{figure}[t!]
    \centering
    \includegraphics[width=\linewidth]{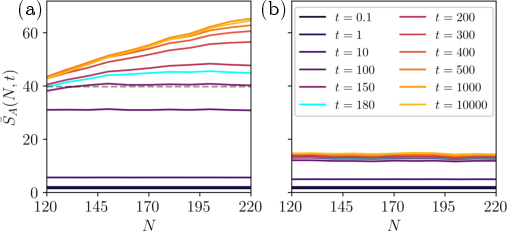}
    \caption{%
        \label{fig:entanglement_entropy}(a, b) Time cuts of the realization-averaged entanglement entropy, $\bar{S}_A$, scaling with the chain length, $N$, for bipartitions $N_A=N/2$ with (a) $W=0.5$ and (b) $W=1.5$.
    }
\end{figure}

As a final investigation into this system, we establish the regime for which the above analysis corresponds to a well-defined topological phase. An SPT phase is defined as being a gapped, short-range entangled phase with a protecting symmetry~\cite{senthil_2015,chiu_2016}. Its EE should, therefore, exhibit area-law behavior~\cite{hastings_2007,eisert_2010,brandao_2013} which in one dimension requires it to be independent of the subsystem partition size, $N_A$. Using this definition, we identify when SPT order becomes ill-defined through a crossover from area- to volume-law scaling, after which the EE will grow (linearly) with $N_A$. This crossover also defines the regime $t>t^*$; when the invariant no longer captures the topological phase of the time-evolved state.

The EE across the subdomain boundary can be determined directly from the corresponding ES via
\begin{align}
    S_A = \sum_\eta \ln(1+e^{-\varepsilon_\eta})+\sum_\eta\frac{\varepsilon_\eta}{e^{\varepsilon_\eta}+1},
\end{align}
where $\varepsilon_\eta=2\arctanh(\xi_\eta)$ are the spectrally flattened ES. Figure~\ref{fig:entanglement_entropy} shows how the realization-averaged EE, $\bar{S}_A$ with $N_A=N/2$, scales with system size for small disorder $W<J^{(f)}$ [Fig.~\ref{fig:entanglement_entropy}~(a)] and large disorder $W>J^{(f)}$ [Fig.~\ref{fig:entanglement_entropy}~(b)]. For small $W$, the EE profile exhibits a crossover from area- to volume-law behavior, signaling a clear breakdown in the SPT order. In addition, this breakdown occurs earlier for chains with smaller $N$ since less time is required for correlations to extend beyond the subsystem partition, which likewise scales with $N$. For example, for a chain of length $N=180$, this crossover can be observed at approximately $t=180$ [cyan curve in Fig.~\ref{fig:entanglement_entropy}~(a)]. This coincides with the time at which the ES gap opens in Fig.~\ref{fig:entanglement_spectra}~(a) and the extended edge modes depicted in Fig.~\ref{fig:localization}~(a) cross the subdomain boundary. However, for large $W$ [Fig.~\ref{fig:entanglement_entropy}~(b)], the EE retains its flat, area-law profile for all $N$. This indicates that the system remains short-range entangled and demonstrates that local disorder can be employed to stabilize SPT order.

%%%%%%%%%%%%%%%%%%%%%%%%%%%%%%%%%%%%%%%%%%%%%%%%%%%%%%%%%%%%%%%%%%%%%%%%%%%%%%%%%%%
%%%%%%%%%%%%%%%%%%%%%%%%%%%%%%%%%%%% DISCUSSION %%%%%%%%%%%%%%%%%%%%%%%%%%%%%%%%%%%
%%%%%%%%%%%%%%%%%%%%%%%%%%%%%%%%%%%%%%%%%%%%%%%%%%%%%%%%%%%%%%%%%%%%%%%%%%%%%%%%%%%

\section{Discussion\label{sec:discussion}}

In this work, we have exposed nontrivial dynamics that depend significantly upon the strength of the local disorder potential. Importantly, for small disorder such that correlations are free to spread throughout the system, we find that the initially localized modes will extend from the edge until a time $t^*$. After this time, SPT order becomes ill-defined and the topological invariant can no longer faithfully describe the system. We identify the $t<t^*$ regime as having a weak breakdown in bulk-boundary correspondence since the MZMs are no longer pinned to the edge but retain an exponential profile and finite correlation length. Our choice to partition the system equidistant from the left and right boundaries thus allows us to identify a maximum time for which SPT order is well-defined since an asymmetric partition would, on average, result in the ES gap opening and the area- to volume-law EE crossover occurring at an earlier time. However, for generic partitions sufficiently far from the edge, the system will maintain a regime featuring a finite correlation length, well-defined topology, and extended edge modes.

Additionally, it is possible to stabilize the topological phase by increasing the value of disorder and suppressing intersite transport. In this regime, the value of $t^*$ increases and the system retains short-range entanglement throughout the time evolution. Yet, we still identify nontrivial dynamics in the polarization and Loschmidt echo corresponding to the preservation of chiral symmetry.

Throughout this study we have considered several distinct quench protocols, which we shall summarize in the following. After initializing the system in a topological phase, we have employed a postquench Hamiltonian that is (i) critical, (ii) Anderson localized, (iii) gapped and topological, or (iv) gapped and nontopological, dictated by tuning disorder, chemical potential and superconducting pairing. For large disorder, where $W$ is the dominant energy scale in the system, the postquench Hamiltonian generally describes an Anderson insulator in which intersite transport is inhibited, resulting in the preservation of the topological phase irrespective of the other quench parameters. However, for low disorder, we find features of the dynamics to be far more varied. For a critical quench ($\Delta^{(f)}=0$), transport renders the system nontopological after sufficient time has elapsed such that correlations can spread throughout the chain, eventually spanning its entirety for systems of finite size. With a nontopological quench, projections onto the eigenstates of the topologically trivial Hamiltonian during the time evolution also render the system nontopological at long time. However, for a gapped and topological quench with $0<\Delta^{(f)}\ll\Delta^{(i)}$, the edge mode partially retains its exponential profile, despite there being only small disorder postquench. This can be attributed to projections onto topological states during time evolution. Further, as the value of superconducting pairing is increased (while keeping disorder small) this effect is enhanced, with the edge states exhibiting increasingly localized profiles at long time with $\Delta^{(f)}\sim\Delta^{(i)}$. Naturally, this coincides with the ES zero-mode gap becoming vanishingly small (with any separation being only due to finite-size effects) and with a complete suppression of oscillations in $\mathrm{CS}_1$ (see \cref{sup:large_delta_quench} for additional information).

In future work, we aim to extend the framework presented here to other noninteracting SPT phases~\cite{verresen_2017}, where the existence of multiple edge Majoranas is possible and we expect more exotic dynamical phase transitions to occur. Beyond this, we aim to study the effect of weak interactions on the dynamics that will naturally couple edge Majoranas as their spatial profiles overlap and may alter the timescales for which we observe a breakdown in SPT order.

%%%%%%%%%%%%%%%%%%%%%%%%%%%%%%%%%%%%%%%%%%%%%%%%%%%%%%%%%%%%%%%%%%%%%%%%%%%%%%%%%%%
%%%%%%%%%%%%%%%%%%%%%%%%%%%%%%%%%%%% END MATTER %%%%%%%%%%%%%%%%%%%%%%%%%%%%%%%%%%%
%%%%%%%%%%%%%%%%%%%%%%%%%%%%%%%%%%%%%%%%%%%%%%%%%%%%%%%%%%%%%%%%%%%%%%%%%%%%%%%%%%%

\begin{acknowledgments}
We thank Teng Ma for insightful discussions. T.L. and M.H. were supported by the National Natural Science Foundation of China (Grant No. 12074039). M.H. was supported by National Natural Science Foundation of China (Grant No. 12150410321). K.P, T.L and M.H. were supported by the Beijing Natural Science Foundation (Grant No. IS24022).\\
\end{acknowledgments}

\appendix

\section{The correlation matrix in the Majorana and fermionic representations}\label{appendix:correlation_matrix}
    Here, we summarize the treatment of the fermion chain using the Majorana representation and the derivation of the correlation matrix. The Hamiltonian can be expressed in terms of the Majorana fermionic operators via the creation and annihilation operators, $c_j^\dagger=\frac{1}{2}(\gamma_{2j}+i\gamma_{2j+1})e^{i\theta/2}$ and $c_j=\frac{1}{2}(\gamma_{2j}-i\gamma_{2j+1})e^{-i\theta/2}$, where $\{\gamma_j, \gamma_k\}=2\delta_{jk}$, $\gamma_j=\gamma_j^\dagger$ and ${(\gamma_j)}^2=1$. By choosing $\theta=0$, the Hamiltonian $H^{(i/f)}$ given in \cref{eq:fermion_ham} of the main text may be expressed as
    \begin{align}
        \label{eq:majorana_ham}
        H^{(\alpha)}=&\frac{i}{2}\sum_j\gamma_{2j}\gamma_{2j+3}(J^{(\alpha)}-|\Delta^{(\alpha)}|)\nn\\
                     &+\gamma_{2j+2}\gamma_{2j+1}(J^{(\alpha)}+|\Delta|^{(\alpha)})+\mu_j^{(\alpha)}\gamma_{2j}\gamma_{2j+1}\nn\\[.5\baselineskip]
        =:&\sum_{ij}\mathcal{H}^{(\alpha)}_{ij}\gamma_i\gamma_j.
    \end{align}
    $H^{(\alpha)}$ is invariant under TRS, PHS and ChS, classifying it as a BDI class system. Thus, there exist unitary matrices $U_T$, $U_C$ and $U_S$, such that $U_T(\mathcal{H}^{(\alpha)})^*U_T^\dagger =\mathcal{H}^{(\alpha)}$, $U_C(\mathcal{H}^{(\alpha)})^*U_C^\dagger =-\mathcal{H}^{(\alpha)}$ and $U_S\mathcal{H}^{(\alpha)}U_S^\dagger =-\mathcal{H}^{(\alpha)}$. In the Majorana representation, these matrices can be expressed as $U_T=\mathbbm{1}_{2N}$ and $U_C=U_S=\mathbbm{1}_N\otimes\sigma_z$, resulting in operators for TRS, PHS and ChS represented by $\mathsf{T}=U_T\mathcal{K}=\mathbbm{1}_{2N}\mathcal{K}$, $\mathsf{C}=\mathbbm{1}_N\otimes\sigma_z\mathcal{K}$ and $\mathsf{S}\equiv U_S=\mathbbm{1}_N\otimes\sigma_z$, respectively, with $\mathcal{K}$ being element-wise complex conjugation. Using these relations, it is straightforward to show that $\mathcal{H}(t)$ respects PHS:
    \begin{align}
        &U_C\mathcal{H}(t)^*U_C\nn\\
        &=U_Ce^{-it\left(\mathcal{H}^{(f)}\right)^*}U_CU_C\left(\mathcal{H}^{(i)}\right)^*U_CU_Ce^{it\left(\mathcal{H}^{(f)}\right)^*}U_C\nn\\
        &=e^{it\mathcal{H}^{(f)}}\left(-\mathcal{H}^{(i)}\right)e^{-it\mathcal{H}^{(f)}} =-\mathcal{H}(t),
    \end{align}
    where we used $U_C\left(\mathcal{H}^{(i/f)}\right)^*U_C =-\mathcal{H}^{(i/f)}$.

    Let us now demonstrate how the correlation matrix in the Majorana representation shown in the main text, $C_{jk}(t)=\avr{\left.\Psi(t)\right|\gamma_j\gamma_k\left|\Psi(t)\right.}$, can be expressed in terms of the instantaneous Majorana eigenfunctions. The single-particle Hamiltonian in the Majorana representation, $\mathcal{H}$, can be diagonalized using its eigenfunctions, $\psi_{m,i}$:
    \begin{align}
    \left(\begin{array}{c} \gamma_{2i} \\ \gamma_{2i+1}\end{array}\right) =& \sum\limits_{m=0}^N \left(\begin{array}{cc} \text{Re}\psi_{m,2i} & \text{Im}\psi_{m,2i} \\ \text{Re}\psi_{m,2i+1} & \text{Im}\psi_{m,2i+1} \end{array}\right) \left(\begin{array}{c} \chi_{2m} \\ \chi_{2m+1}\end{array}\right),
    \end{align}
    with the new set of Majorana operators $\chi_{2m}$, $\chi_{2m+1}$. The particle-hole symmetry of the system is reflected in the properties of the Hamiltonian in the following way: the matrix $i\mathcal{H}_{ij}$ is both antisymmetric and real, so for every eigenfunction $\psi_{m,i}$ with energy $E_m$ corresponds $\psi^*_{m,i}$ with energy $-E_m$. Diagonalizing the Hamiltonian with the above transformation, we find that
    \begin{align}
    H =&\sum\limits_{i,j=0}^{2N}\mathcal{H}_{ij}\gamma_i\gamma_j =\sum\limits_{m=0}^{N}E_m\frac{-i}{2}\chi_{2m}\chi_{2m+1} \\
    =&\sum\limits_{m=0}^N E_m \left(d_m^\dagger d_m -\frac{1}{2}\right),
    \end{align}
    where $d_m^{(\dagger)}$ is the annihilation (creation) operator for the $m$'th energy eigenstate in the fermionic representation.

    Suppose we prepare the system in the state of $n-$filling, i.e. $\ket{\Psi_n} =\prod_{m=0}^n \gamma_m \ket{0}$. The correlation matrix can be written as
    \begin{align}
    C_{ij,n} =& \bra{\Psi_n}\gamma_i\gamma_j\ket{\Psi_n} \\
    = \sum_{mm'}\bra{\Psi_n}&\left(\text{Re}\psi_{m,i}\chi_{2m} +\text{Im}\psi_{m,i}\chi_{2m+1}\right) \nn\\
    & \times\left(\text{Re}\psi_{m',j}\chi_{2m'} +\text{Im}\psi_{m',j}\chi_{2m'+1}\right)\ket{\Psi_n}.
    \end{align}
    Utilizing that $-i\chi_{2m}\chi_{2m+1}$ gives $\pm 1$ for $m\leq n$ ($m>n$):
    \begin{align}
    C_{ij,n} =\sum_m & \left[\text{Re}\psi_{m,i}\text{Re}\psi_{m,j} +\text{Im}\psi_{m, i}\text{Im}\psi_{m, j} \right. \nn\\
    & +i\left(\text{Re}\psi_{m,i}\text{Im}\psi_{m,j}-\text{Im}\psi_{m,i}\text{Re}\psi_{m,j}\right)\nn\\
    &\times\left. \left(\theta(n-m) -\theta(m-n)\right) \right],
    \end{align}
    which can be expressed as
    \begin{align}
    C_{ij,n} =& \delta_{ij} +2i\text{Im}\sum_{m\leq n}\psi^*_{m,i}\psi_{m,j}.
    \end{align}
    Specifically, for half-filling (i.e., $n=N$) one can utilize the relation $\sum_{m=0}^N\text{Re}(\psi^*_{m,i}\psi_{m,j}) =\frac{1}{2}\delta_{ij}$, which holds because of $(i)$ the completeness of the eigenfunctions as a basis and $(ii)$ particle-hole symmetry, to realize that $$C_{ij} =2\sum\limits_{m=0}^N\psi^*_{m,i}\psi_{m,j},$$ which we have also quoted in the main text. For half-filling, we omit the index $n$ throughout the main text. One can summarize the structure of the Majorana correlation matrix as $C_{ij}=\avr{\gamma_i\gamma_j} =\delta_{ij} +i\Gamma_{ij}$, where $\Gamma_{ij}$ is real and antisymmetric.

    Let us now discuss how the correlations matrix, $C$, is expressed in the fermionic representation by using the matrices $\mathcal{C}=\avr{c^\dagger c}$ and $\mathcal{F}=\avr{cc}$. To establish these relations we first separate the odd and even parity indices, then utilize the following transformation:
    \begin{align}
    \left(\begin{array}{c} c_i \\ c_i^\dagger \end{array}\right) =& \frac{1}{2}\left(\begin{array}{cc} 1 & -i \\ 1 & i \end{array}\right)\left(\begin{array}{c} \gamma_{2i} \\ \gamma_{2i+1} \end{array}\right).
    \end{align}
    This allows us to express the components of $C_{ij}$ in terms of their fermionic counterpart,
    \begin{align}
    & \frac{1}{2}\left(\begin{array}{cc} 1 & -i \\ 1 & i \end{array}\right) \underbrace{\left(\begin{array}{cc} \avr{\gamma_{2i}\gamma_{2j}} & \avr{\gamma_{2i}\gamma_{2j+1}} \\ \avr{\gamma_{2i+1}\gamma_{2j}} & \avr{\gamma_{2i+1}\gamma_{2j+1}} \end{array}\right)}_{=:\widehat{C}_{ij}}\left(\begin{array}{cc} 1 & 1 \\ i & -i \end{array}\right) \nn\\
    & =2\left(\begin{array}{cc} \avr{c_ic\dagger_j} & \avr{c_ic_j} \\ \avr{c^\dagger_ic^\dagger_j} & \avr{c^\dagger_ic_j} \end{array}\right) =2\left(\begin{array}{cc} \delta_{ij}-\mathcal{C}_{ji}& \mathcal{F}_{ij} \\ \mathcal{F}^*_{ji} & \mathcal{C}_{ij}\end{array}\right), \label{eq:changeRepC}
    \end{align}
    where we have employed the shorthand $\avr{\dots}\equiv \bra{\Psi(t)}\dots\ket{\Psi(t)}$. Indices with different parity involve different combinations of the correlation matrices $\mathcal{C}$ and $\mathcal{F}$. These correlation matrices have the following properties with respect to interchanging indices:
        \begin{align}
            \mathcal{F}_{ij}(t) =& \avr{c_i(t)c_j(t)}=-\avr{c_j(t)c_i(t)} =-\mathcal{F}_{ji}(t),\nn\\
            \mathcal{F}^*_{ij}(t) =& \avr{c^\dagger_j(t)c^\dagger_i(t)} =-\avr{c^\dagger_i(t)c^\dagger_j(t)} =-\mathcal{F}^*_{ji}(t),\\
            \mathcal{C}^*_{ij}(t) =& \avr{c^\dagger_j(t)c_i(t)} =\mathcal{C}_{ji}(t). \nn
        \end{align}

    In the $N\rightarrow\infty$ limit, the midgap energy eigenstates of $\mathcal{H}^{(i/f)}$ are exactly degenerate in the topological regime. Thus, distinguishing which of the modes is populated at half-filling is ambiguous. To resolve this ambiguity, we take a linear combination of these states. For the system size simulated in this work, the states are sufficiently close to degenerate that this procedure is also applicable. Therefore, we construct a new state in the zero-energy subspace as a general linear combination of $\ket{\psi_-}$ and $\ket{\psi_+}$: the ``lower'' and ``upper'' zero modes, respectively. This takes the form $\ket{\theta}=\cos(\theta)\ket{\psi_-}+\sin(\theta)\ket{\psi_+}$, where $\theta$ is an arbitrary phase. Elements of the correlation matrix in the zero-energy subspace are then given by the dyadic product, $C^0_{jk}(\theta)=\ket{\theta_j}\bra{\theta_k}$. To account for all possible choices of $\theta$ we integrate over this manifold with an equal weight, $\frac{1}{2\pi}\int_0^{2\pi}d\theta C^0_{jk}=\frac{1}{2}(\ket{\psi_{-,j}}\bra{\psi_{-,k}}+\ket{\psi_{+,j}}\bra{\psi_{+,k}})$. As a consequence, cross-terms between $\ket{\psi_-}$ and $\ket{\psi_+}$ vanish, and we can include the zero modes in the correlation matrix of states up to half filling by simply summing over them (as in the definition for $C_{jk}$ in the main text) with an additional prefactor of $1/\sqrt{2}$.\\

\section{Dynamics of the pseudospin}\label{sup:psSpinDynamics}
    \subsection{Pseudospin equation of motion for an inhomogeneous system}
    We briefly look into the pseudospin dynamics of an inhomogeneous system --- in the infinite system size limit --- to capture the dynamics of the disordered case. This is described by the postquench Hamiltonian $H^{(f)}=\sum_{Xkl}e^{-iXl}\Phi^\dagger_{k+\frac{l}{2}}\bm{\mathcal{D}}(X,k)\cdot\bm{\sigma}\Phi_{k-\frac{l}{2}}$, with $\bm{\mathcal{D}}(X,k)=\ve{d}^{(f)}(k) +(0,\,0,\,-\mathcal{W}(X))$, where $\mathcal{W}(x)$ is the disorder potential in a given realization of the system. The equation of motion for the averaged pseudospin operator is thus given by $\ve{s}(k,t)=\sum_{Xy}e^{iky}\Phi^\dagger_{X+\frac{y}{2}}\bm{\sigma}\Phi_{X-\frac{y}{2}}$. In terms of the correlation matrix in the Majorana representation, the expectation value for the pseudospin texture can then be written as,
    \begin{widetext}
    \begin{align}
    \partial_t \ve{n}^{(M)}(k,t)
    =& 2\ve{d}^{(M,f)}(k)\times\ve{n}^{(M)}(k,t) -\frac{1}{N^2}\sum_{jj'}e^{ik(j-j')}(\mathcal{W}_j-\mathcal{W}_{j'})\frac{1}{2}\text{tr}\left\{ \left(\sigma_z,\,i\mathbbm{1},\,-\sigma_x\right)\widehat{C}_{jj'}(t) \right\},
    \end{align}\end{widetext}
    where the elements of $\ve{d}^{(M,f)}(k)$ simply come from a permutation of the elements of $\ve{d}^{(f)}(k)$. Beyond the pseudospin-precession observed in a clean system, there are now additional terms which require the evaluation of not only the pseudospin vector, but components of the correlation matrix itself --- weighted by the value of the disorder potential $\mathcal{W}_j$. The D class index, $\nu_D(t) =\frac{1}{2}\left(\text{sign}(n^{(M)}_y(k=\pi,t))-\text{sign}(n^{(M)}_y(k=0,t))\right)$, corresponding to a given realization now depends not only on the initial pseudospin configuration, but carries the history of the correlation matrix as well, since for the $y-$component of the pseudospin-texture in the Majorana representation:
    \begin{widetext}
    \begin{align}
    \left(\begin{array}{c} n^{(M)}_y(k=0,t) \\ n^{(M)}_y(k=\pi,t) \end{array}\right)
    =& \left(\begin{array}{c} n^{(M,i)}_y(k=0) \\ n^{(M,i)}_y(k=\pi) \end{array}\right) +\frac{1}{2N^2}\sum_{jj'}\left(\begin{array}{c} 1 \\ (-1)^{j-j'}\end{array}\right) (\mathcal{W}_{j}-\mathcal{W}_{j'})\text{tr}\left\{\intlim{\tau}{0}{t} \widehat{\Gamma}_{jj'}(\tau)\right\}.
    \end{align}
    \end{widetext}

    \subsection{Pseudospin from the correlation matrix}\label{sup:psPsinCorrMx}
    Here, we outline how to express $\ve{n}$ in terms of the single-particle correlation matrix. To derive \cref{eq:psSpinX,eq:psSpinY,eq:psSpinZ}, we write the expectation value of the pseudospin operator in terms of $\avr{\wt{c}^{(\dagger)}_k(t)\wt{c}^{(\dagger)}_{-k}(t)}$ and $\avr{\wt{c}^\dagger_k(t)\wt{c}_k(t)}$, then express them in terms of the Majorana correlation matrix by using \cref{eq:changeRepC}. Through straightforward algebraic steps and manipulating the indices $i$ and $j$, we arrive at the manifestly real-valued expressions:
        \begin{widetext}
            \begin{align}
                n_x(k,t)=&
                -\avr{\Phi^\dagger_k(t)\sigma_x\Phi_k(t)}
                =-\frac{1}{N}\sum_{ij}e^{-ik(x_i-x_j)}\left( \avr{c^\dagger_i(t)c^\dagger_j(t)} +\avr{c_i(t)c_j(t)} \right) \nn\\
                =& -\frac{1}{N}\sum_{ij}\sin(k(x_i-x_j))\frac{\mathcal{F}_{ij}(t)-\mathcal{F}^*_{ij}(t)}{i}
                =\frac{1}{N}\sum_{ij}\sin(k(x_i-x_j))\frac{\Gamma_{2i+1,2j+1}(t) -\Gamma_{2i,2j}(t)}{2}, \label{eq:psSpinCorrMx1}
            \end{align}
            \begin{align}
                n_y(k,t)=&
                -\avr{\Phi^\dagger_k(t)\sigma_y\Phi_k(t)}
                = -\frac{1}{N}\sum_{ij}e^{-ik(x_i-x_j)}(-i)\left( \avr{c^\dagger_i(t)c^\dagger_j(t)} -\avr{c_i(t)c_j(t)} \right) \nn\\
                =& -\frac{1}{N}\sum_{ij}\sin(k(x_i-x_j)) \left( \mathcal{F}_{ij}(t)+\mathcal{F}^*_{ij}(t) \right)
                =-\frac{1}{N}\sum_{ij}\sin(k(x_i-x_j))\frac{\Gamma_{2i+1,2j}(t) +\Gamma_{2i,2j+1}(t)}{2} , \label{eq:psSpinCorrMx2}
            \end{align}
            \begin{align}
                n_z(k,t)=&
                -\avr{\Phi^\dagger_k(t)\sigma_z\Phi_k(t)}
                =-\frac{1}{N}\sum_{ij}e^{-ik(x_i-x_j)}\left( \avr{c^\dagger_i(t)c_j(t)} -\avr{c_i(t)c^\dagger_j(t)} \right) \nn\\
                =& -\frac{1}{N}\sum_{i\neq j}\cos(k(x_i-x_j))\left(\mathcal{C}_{ij}(t)+\mathcal{C}_{ji}(t)\right) -\frac{1}{N}\sum_i\avr{c_i^\dagger(t)c_i(t)-c_i(t)c_i^\dagger(t)} \nn\\
                = &1- \frac{1}{N}\sum_{ij}\cos(k(x_i-x_j))\left(\mathcal{C}_{ij}(t)+\mathcal{C}_{ij}^*(t)\right)
                =\frac{1}{N}\sum_{ij}\cos(k(x_i-x_j))\frac{\Gamma_{2i+1,2j}(t) -\Gamma_{2i,2j+1}(t)}{2}, \label{eq:psSpinCorrMx3}
            \end{align}
        \end{widetext}
        where in the last step the expressions for the pseudospin components are given in terms of the Majorana correlation matrix $\Gamma$. The structure of \cref{eq:psSpinCorrMx1,eq:psSpinCorrMx2,eq:psSpinCorrMx3} is determined by PHS through the form of the spinors. Utilizing the antisymmetry of $\Gamma_{ij}$ in its indices, one can conclude that $n_x(-k,t)=-n_x(k,t)$, $n_y(-k,t)=-n_y(k,t)$ and $n_z(-k,t)=n_z(k,t)$. This can be observed by interchanging the indices $i$ and $j$ in the expression for the components of $\ve{n}(-k,t)$ and identifying $\ve{n}(k,t)$ in the result, up to a sign. Note, that for periodic systems the above is equivalent to taking the Fourier transform of the correlation matrix. The interested reader can find a treatment similar to the one presented above in Ref.~\cite{hsu_2021}.

\section{Quench data for \texorpdfstring{$\Delta^{(f)}=\Delta^{(i)}=0.9$}{large superconducting pairing}}\label{sup:large_delta_quench}

\begin{figure*}[!ht]
    \centering
    \includegraphics[width=\textwidth]{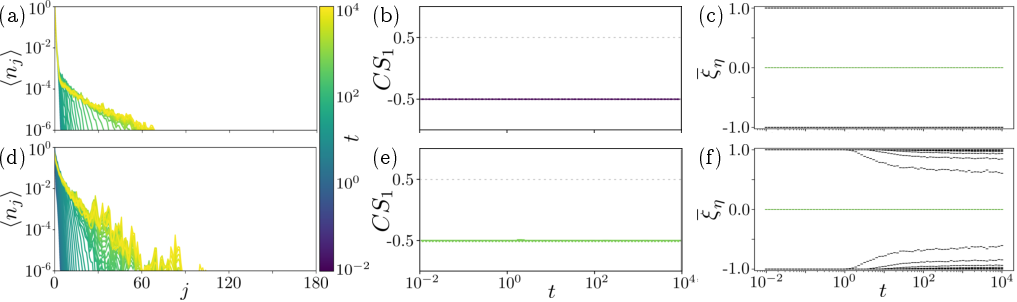}
    \caption{%
        \label{fig:finite_delta_appendix}Time evolution of various properties for a chain of length $N=180$, initialized in the same initial topological phase as in Fig.~\ref{fig:localization} and with a quench defined by $(J^{(f)},~\Delta^{(f)},~\mu_j^{(f)})=(J^{(i)},~0.9,~0.2+w_j)$ with disorder (a-c) $W=0.2$ and (d-f) $W=5.0$. (a, d) Realization-averaged population profiles $\braket{n_{j}}$ for the MZM initially localized at the left edge of a chain with OBC. (b, e) Individual realizations of $\mathrm{CS}_1$, calculated for equivalent chains to those in panels (a, d), respectively, but with PBC. (c, f) Realization-averaged entanglement spectrum for the same chains as in panels (a, d), respectively.
    }
\end{figure*}

In the following, we briefly introduce the dynamics observed for quenches featuring superconducting pairing equal to that of the initial system, $\Delta^{(f)}=\Delta^{(i)}=0.9$, for both small ($W\ll J^{(f)},\Delta^{(f)}$) and large ($W\gg J^{(f)},\Delta^{(f)}$) disorder windows. Continuing the trend seen in Fig.~\ref{fig:localization} and discussed in \cref{sec:discussion}, the localization profiles at long time for quenches with large $\Delta^{(f)}$ retain significantly more of the initial exponential localization when compared to those for smaller $\Delta^{(f)}$. However, in contrast with the results depicted previously in Fig.~\ref{fig:localization}, for $\Delta^{(f)}=0.9$ the profile for high disorder [Fig.~\ref{fig:finite_delta_appendix}~(d)] now appears to be less strongly localized than for low disorder [Fig.~\ref{fig:finite_delta_appendix}~(a)]. while at first glance this might seem somewhat counter-intuitive, it can be understood by considering the tunneling pathways available between Majoranas. When the pairing approaches the value of the intersite hopping, terms appearing as $J-|\Delta|$ in \cref{eq:majorana_ham} will vanish. The postquench Hamiltonian then reduces to a chain of $2N$ Majoranas with alternating hopping amplitudes $J+|\Delta|$ and $\mu_j$. Consequentially, increasing disorder (and thus $\mu_j$) enhances transport between odd/even Majorana subspecies and results in an enhancement in propagation away from the edge. Despite this distinction, however, neither profile extends significantly into the bulk and as such both retain the gapless, degenerate zero-modes in the profiles of their entanglement spectra. This would indicate that large disorder is not required to stabilize the SPT phase when the system features large superconducting pairing.

Interestingly, large $\Delta^{(f)}$ has the additional effect of suppressing oscillations in $\mathrm{CS}_1$ [see panels (b) and (e) of Fig.~\ref{fig:finite_delta_appendix}], though we do find that greater values of $\Delta^{(f)}$ are required to suppress oscillations when in the presence of larger disorder. Further, it also reduces the density of the $k-$space grid required to cleanly resolve $\mathrm{CS}_1$ as compared to Figs.~\ref{fig:localization}~(b) and \ref{fig:localization}~(d) of the main text.

\bibliography{extended_edge.bib}

%apsrev4-2.bst 2019-01-14 (MD) hand-edited version of apsrev4-1.bst
%Control: key (0)
%Control: author (8) initials jnrlst
%Control: editor formatted (1) identically to author
%Control: production of article title (0) allowed
%Control: page (0) single
%Control: year (1) truncated
%Control: production of eprint (0) enabled
\begin{thebibliography}{65}%
\makeatletter
\providecommand \@ifxundefined [1]{%
 \@ifx{#1\undefined}
}%
\providecommand \@ifnum [1]{%
 \ifnum #1\expandafter \@firstoftwo
 \else \expandafter \@secondoftwo
 \fi
}%
\providecommand \@ifx [1]{%
 \ifx #1\expandafter \@firstoftwo
 \else \expandafter \@secondoftwo
 \fi
}%
\providecommand \natexlab [1]{#1}%
\providecommand \enquote  [1]{``#1''}%
\providecommand \bibnamefont  [1]{#1}%
\providecommand \bibfnamefont [1]{#1}%
\providecommand \citenamefont [1]{#1}%
\providecommand \href@noop [0]{\@secondoftwo}%
\providecommand \href [0]{\begingroup \@sanitize@url \@href}%
\providecommand \@href[1]{\@@startlink{#1}\@@href}%
\providecommand \@@href[1]{\endgroup#1\@@endlink}%
\providecommand \@sanitize@url [0]{\catcode `\\12\catcode `\$12\catcode `\&12\catcode `\#12\catcode `\^12\catcode `\_12\catcode `\%12\relax}%
\providecommand \@@startlink[1]{}%
\providecommand \@@endlink[0]{}%
\providecommand \url  [0]{\begingroup\@sanitize@url \@url }%
\providecommand \@url [1]{\endgroup\@href {#1}{\urlprefix }}%
\providecommand \urlprefix  [0]{URL }%
\providecommand \Eprint [0]{\href }%
\providecommand \doibase [0]{https://doi.org/}%
\providecommand \selectlanguage [0]{\@gobble}%
\providecommand \bibinfo  [0]{\@secondoftwo}%
\providecommand \bibfield  [0]{\@secondoftwo}%
\providecommand \translation [1]{[#1]}%
\providecommand \BibitemOpen [0]{}%
\providecommand \bibitemStop [0]{}%
\providecommand \bibitemNoStop [0]{.\EOS\space}%
\providecommand \EOS [0]{\spacefactor3000\relax}%
\providecommand \BibitemShut  [1]{\csname bibitem#1\endcsname}%
\let\auto@bib@innerbib\@empty
%</preamble>
\bibitem [{\citenamefont {Altland}\ and\ \citenamefont {Zirnbauer}(1997)}]{altland_1997}%
  \BibitemOpen
  \bibfield  {author} {\bibinfo {author} {\bibfnamefont {A.}~\bibnamefont {Altland}}\ and\ \bibinfo {author} {\bibfnamefont {M.~R.}\ \bibnamefont {Zirnbauer}},\ }\bibfield  {title} {\bibinfo {title} {Nonstandard symmetry classes in mesoscopic normal-superconducting hybrid structures},\ }\href {https://doi.org/10.1103/PhysRevB.55.1142} {\bibfield  {journal} {\bibinfo  {journal} {Phys. Rev. B}\ }\textbf {\bibinfo {volume} {55}},\ \bibinfo {pages} {1142} (\bibinfo {year} {1997})}\BibitemShut {NoStop}%
\bibitem [{\citenamefont {Kitaev}(2009)}]{kitaev_2009}%
  \BibitemOpen
  \bibfield  {author} {\bibinfo {author} {\bibfnamefont {A.}~\bibnamefont {Kitaev}},\ }\bibfield  {title} {\bibinfo {title} {Periodic table for topological insulators and superconductors},\ }\href {https://doi.org/10.1063/1.3149495} {\bibfield  {journal} {\bibinfo  {journal} {AIP Conf. Proc.}\ }\textbf {\bibinfo {volume} {1134}},\ \bibinfo {pages} {22} (\bibinfo {year} {2009})}\BibitemShut {NoStop}%
\bibitem [{\citenamefont {Ryu}\ \emph {et~al.}(2010)\citenamefont {Ryu}, \citenamefont {Schnyder}, \citenamefont {Furusaki},\ and\ \citenamefont {Ludwig}}]{ryu_2010}%
  \BibitemOpen
  \bibfield  {author} {\bibinfo {author} {\bibfnamefont {S.}~\bibnamefont {Ryu}}, \bibinfo {author} {\bibfnamefont {A.~P.}\ \bibnamefont {Schnyder}}, \bibinfo {author} {\bibfnamefont {A.}~\bibnamefont {Furusaki}},\ and\ \bibinfo {author} {\bibfnamefont {A.~W.~W.}\ \bibnamefont {Ludwig}},\ }\bibfield  {title} {\bibinfo {title} {Topological insulators and superconductors: tenfold way and dimensional hierarchy},\ }\href {https://doi.org/10.1088/1367-2630/12/6/065010} {\bibfield  {journal} {\bibinfo  {journal} {New J. Phys.}\ }\textbf {\bibinfo {volume} {12}},\ \bibinfo {pages} {065010} (\bibinfo {year} {2010})}\BibitemShut {NoStop}%
\bibitem [{\citenamefont {Chiu}\ \emph {et~al.}(2016)\citenamefont {Chiu}, \citenamefont {Teo}, \citenamefont {Schnyder},\ and\ \citenamefont {Ryu}}]{chiu_2016}%
  \BibitemOpen
  \bibfield  {author} {\bibinfo {author} {\bibfnamefont {C.-K.}\ \bibnamefont {Chiu}}, \bibinfo {author} {\bibfnamefont {J.~C.~Y.}\ \bibnamefont {Teo}}, \bibinfo {author} {\bibfnamefont {A.~P.}\ \bibnamefont {Schnyder}},\ and\ \bibinfo {author} {\bibfnamefont {S.}~\bibnamefont {Ryu}},\ }\bibfield  {title} {\bibinfo {title} {Classification of topological quantum matter with symmetries},\ }\href {https://doi.org/10.1103/RevModPhys.88.035005} {\bibfield  {journal} {\bibinfo  {journal} {Rev. Mod. Phys.}\ }\textbf {\bibinfo {volume} {88}},\ \bibinfo {pages} {035005} (\bibinfo {year} {2016})}\BibitemShut {NoStop}%
\bibitem [{\citenamefont {McGinley}\ and\ \citenamefont {Cooper}(2019)}]{mcginley_2019}%
  \BibitemOpen
  \bibfield  {author} {\bibinfo {author} {\bibfnamefont {M.}~\bibnamefont {McGinley}}\ and\ \bibinfo {author} {\bibfnamefont {N.~R.}\ \bibnamefont {Cooper}},\ }\bibfield  {title} {\bibinfo {title} {Classification of topological insulators and superconductors out of equilibrium},\ }\href {https://doi.org/10.1103/PhysRevB.99.075148} {\bibfield  {journal} {\bibinfo  {journal} {Phys. Rev. B}\ }\textbf {\bibinfo {volume} {99}},\ \bibinfo {pages} {075148} (\bibinfo {year} {2019})}\BibitemShut {NoStop}%
\bibitem [{\citenamefont {Chung}\ \emph {et~al.}(2013)\citenamefont {Chung}, \citenamefont {Jhu}, \citenamefont {Chen},\ and\ \citenamefont {Mou}}]{chung_2013}%
  \BibitemOpen
  \bibfield  {author} {\bibinfo {author} {\bibfnamefont {M.-C.}\ \bibnamefont {Chung}}, \bibinfo {author} {\bibfnamefont {Y.-H.}\ \bibnamefont {Jhu}}, \bibinfo {author} {\bibfnamefont {P.}~\bibnamefont {Chen}},\ and\ \bibinfo {author} {\bibfnamefont {C.-Y.}\ \bibnamefont {Mou}},\ }\bibfield  {title} {\bibinfo {title} {Quench dynamics of topological maximally entangled states},\ }\href {https://doi.org/10.1088/0953-8984/25/28/285601} {\bibfield  {journal} {\bibinfo  {journal} {J. Phys.: Condens. Matter}\ }\textbf {\bibinfo {volume} {25}},\ \bibinfo {pages} {285601} (\bibinfo {year} {2013})}\BibitemShut {NoStop}%
\bibitem [{\citenamefont {Hegde}\ \emph {et~al.}(2015)\citenamefont {Hegde}, \citenamefont {Shivamoggi}, \citenamefont {Vishveshwara},\ and\ \citenamefont {Sen}}]{hegde_2015}%
  \BibitemOpen
  \bibfield  {author} {\bibinfo {author} {\bibfnamefont {S.}~\bibnamefont {Hegde}}, \bibinfo {author} {\bibfnamefont {V.}~\bibnamefont {Shivamoggi}}, \bibinfo {author} {\bibfnamefont {S.}~\bibnamefont {Vishveshwara}},\ and\ \bibinfo {author} {\bibfnamefont {D.}~\bibnamefont {Sen}},\ }\bibfield  {title} {\bibinfo {title} {Quench dynamics and parity blocking in majorana wires},\ }\href {https://doi.org/10.1088/1367-2630/17/5/053036} {\bibfield  {journal} {\bibinfo  {journal} {New J. Phys.}\ }\textbf {\bibinfo {volume} {17}},\ \bibinfo {pages} {053036} (\bibinfo {year} {2015})}\BibitemShut {NoStop}%
\bibitem [{\citenamefont {Hegde}\ and\ \citenamefont {Vishveshwara}(2016)}]{hegde_2016}%
  \BibitemOpen
  \bibfield  {author} {\bibinfo {author} {\bibfnamefont {S.~S.}\ \bibnamefont {Hegde}}\ and\ \bibinfo {author} {\bibfnamefont {S.}~\bibnamefont {Vishveshwara}},\ }\bibfield  {title} {\bibinfo {title} {Majorana wave-function oscillations, fermion parity switches, and disorder in kitaev chains},\ }\href {https://doi.org/10.1103/PhysRevB.94.115166} {\bibfield  {journal} {\bibinfo  {journal} {Phys. Rev. B}\ }\textbf {\bibinfo {volume} {94}},\ \bibinfo {pages} {115166} (\bibinfo {year} {2016})}\BibitemShut {NoStop}%
\bibitem [{\citenamefont {Chung}\ \emph {et~al.}(2016)\citenamefont {Chung}, \citenamefont {Jhu}, \citenamefont {Chen}, \citenamefont {Mou},\ and\ \citenamefont {Wan}}]{chung_2016}%
  \BibitemOpen
  \bibfield  {author} {\bibinfo {author} {\bibfnamefont {M.-C.}\ \bibnamefont {Chung}}, \bibinfo {author} {\bibfnamefont {Y.-H.}\ \bibnamefont {Jhu}}, \bibinfo {author} {\bibfnamefont {P.}~\bibnamefont {Chen}}, \bibinfo {author} {\bibfnamefont {C.-Y.}\ \bibnamefont {Mou}},\ and\ \bibinfo {author} {\bibfnamefont {X.}~\bibnamefont {Wan}},\ }\bibfield  {title} {\bibinfo {title} {A memory of majorana modes through quantum quench},\ }\href {https://doi.org/10.1038/srep29172} {\bibfield  {journal} {\bibinfo  {journal} {Sci. Rep.}\ }\textbf {\bibinfo {volume} {6}},\ \bibinfo {pages} {29172} (\bibinfo {year} {2016})}\BibitemShut {NoStop}%
\bibitem [{\citenamefont {McGinley}\ and\ \citenamefont {Cooper}(2018)}]{mcginley_2018}%
  \BibitemOpen
  \bibfield  {author} {\bibinfo {author} {\bibfnamefont {M.}~\bibnamefont {McGinley}}\ and\ \bibinfo {author} {\bibfnamefont {N.~R.}\ \bibnamefont {Cooper}},\ }\bibfield  {title} {\bibinfo {title} {Topology of one-dimensional quantum systems out of equilibrium},\ }\href {https://doi.org/10.1103/PhysRevLett.121.090401} {\bibfield  {journal} {\bibinfo  {journal} {Phys. Rev. Lett.}\ }\textbf {\bibinfo {volume} {121}},\ \bibinfo {pages} {090401} (\bibinfo {year} {2018})}\BibitemShut {NoStop}%
\bibitem [{\citenamefont {Qi}\ \emph {et~al.}(2008)\citenamefont {Qi}, \citenamefont {Hughes},\ and\ \citenamefont {Zhang}}]{qi_2008}%
  \BibitemOpen
  \bibfield  {author} {\bibinfo {author} {\bibfnamefont {X.-L.}\ \bibnamefont {Qi}}, \bibinfo {author} {\bibfnamefont {T.~L.}\ \bibnamefont {Hughes}},\ and\ \bibinfo {author} {\bibfnamefont {S.-C.}\ \bibnamefont {Zhang}},\ }\bibfield  {title} {\bibinfo {title} {Topological field theory of time-reversal invariant insulators},\ }\href {https://doi.org/10.1103/PhysRevB.78.195424} {\bibfield  {journal} {\bibinfo  {journal} {Phys. Rev. B}\ }\textbf {\bibinfo {volume} {78}},\ \bibinfo {pages} {195424} (\bibinfo {year} {2008})}\BibitemShut {NoStop}%
\bibitem [{\citenamefont {Gong}\ and\ \citenamefont {Ueda}(2018)}]{gong_2018}%
  \BibitemOpen
  \bibfield  {author} {\bibinfo {author} {\bibfnamefont {Z.}~\bibnamefont {Gong}}\ and\ \bibinfo {author} {\bibfnamefont {M.}~\bibnamefont {Ueda}},\ }\bibfield  {title} {\bibinfo {title} {Topological entanglement-spectrum crossing in quench dynamics},\ }\href {https://doi.org/10.1103/PhysRevLett.121.250601} {\bibfield  {journal} {\bibinfo  {journal} {Phys. Rev. Lett.}\ }\textbf {\bibinfo {volume} {121}},\ \bibinfo {pages} {250601} (\bibinfo {year} {2018})}\BibitemShut {NoStop}%
\bibitem [{\citenamefont {Sedlmayr}\ \emph {et~al.}(2018)\citenamefont {Sedlmayr}, \citenamefont {Jaeger}, \citenamefont {Maiti},\ and\ \citenamefont {Sirker}}]{sedlmayr_2018}%
  \BibitemOpen
  \bibfield  {author} {\bibinfo {author} {\bibfnamefont {N.}~\bibnamefont {Sedlmayr}}, \bibinfo {author} {\bibfnamefont {P.}~\bibnamefont {Jaeger}}, \bibinfo {author} {\bibfnamefont {M.}~\bibnamefont {Maiti}},\ and\ \bibinfo {author} {\bibfnamefont {J.}~\bibnamefont {Sirker}},\ }\bibfield  {title} {\bibinfo {title} {Bulk-boundary correspondence for dynamical phase transitions in one-dimensional topological insulators and superconductors},\ }\href {https://doi.org/10.1103/PhysRevB.97.064304} {\bibfield  {journal} {\bibinfo  {journal} {Phys. Rev. B}\ }\textbf {\bibinfo {volume} {97}},\ \bibinfo {pages} {064304} (\bibinfo {year} {2018})}\BibitemShut {NoStop}%
\bibitem [{\citenamefont {Pastori}\ \emph {et~al.}(2020)\citenamefont {Pastori}, \citenamefont {Barbarino},\ and\ \citenamefont {Budich}}]{pastori_2020}%
  \BibitemOpen
  \bibfield  {author} {\bibinfo {author} {\bibfnamefont {L.}~\bibnamefont {Pastori}}, \bibinfo {author} {\bibfnamefont {S.}~\bibnamefont {Barbarino}},\ and\ \bibinfo {author} {\bibfnamefont {J.~C.}\ \bibnamefont {Budich}},\ }\bibfield  {title} {\bibinfo {title} {Signatures of topology in quantum quench dynamics and their interrelation},\ }\href {https://doi.org/10.1103/PhysRevResearch.2.033259} {\bibfield  {journal} {\bibinfo  {journal} {Phys. Rev. Res.}\ }\textbf {\bibinfo {volume} {2}},\ \bibinfo {pages} {033259} (\bibinfo {year} {2020})}\BibitemShut {NoStop}%
\bibitem [{\citenamefont {Sayyad}\ \emph {et~al.}(2021)\citenamefont {Sayyad}, \citenamefont {Yu}, \citenamefont {Grushin},\ and\ \citenamefont {Sieberer}}]{sayyad_2021}%
  \BibitemOpen
  \bibfield  {author} {\bibinfo {author} {\bibfnamefont {S.}~\bibnamefont {Sayyad}}, \bibinfo {author} {\bibfnamefont {J.}~\bibnamefont {Yu}}, \bibinfo {author} {\bibfnamefont {A.~G.}\ \bibnamefont {Grushin}},\ and\ \bibinfo {author} {\bibfnamefont {L.~M.}\ \bibnamefont {Sieberer}},\ }\bibfield  {title} {\bibinfo {title} {Entanglement spectrum crossings reveal non-hermitian dynamical topology},\ }\href {https://doi.org/10.1103/PhysRevResearch.3.033022} {\bibfield  {journal} {\bibinfo  {journal} {Phys. Rev. Res.}\ }\textbf {\bibinfo {volume} {3}},\ \bibinfo {pages} {033022} (\bibinfo {year} {2021})}\BibitemShut {NoStop}%
\bibitem [{\citenamefont {Zhang}\ \emph {et~al.}(2022)\citenamefont {Zhang}, \citenamefont {Jia},\ and\ \citenamefont {Liu}}]{zhang_2022}%
  \BibitemOpen
  \bibfield  {author} {\bibinfo {author} {\bibfnamefont {L.}~\bibnamefont {Zhang}}, \bibinfo {author} {\bibfnamefont {W.}~\bibnamefont {Jia}},\ and\ \bibinfo {author} {\bibfnamefont {X.-J.}\ \bibnamefont {Liu}},\ }\bibfield  {title} {\bibinfo {title} {Universal topological quench dynamics for z2 topological phases},\ }\href {https://doi.org/https://doi.org/10.1016/j.scib.2022.04.019} {\bibfield  {journal} {\bibinfo  {journal} {Sci. Bull.}\ }\textbf {\bibinfo {volume} {67}},\ \bibinfo {pages} {1236} (\bibinfo {year} {2022})}\BibitemShut {NoStop}%
\bibitem [{\citenamefont {Fang}\ \emph {et~al.}(2022)\citenamefont {Fang}, \citenamefont {Wang},\ and\ \citenamefont {Li}}]{fang_2022}%
  \BibitemOpen
  \bibfield  {author} {\bibinfo {author} {\bibfnamefont {P.}~\bibnamefont {Fang}}, \bibinfo {author} {\bibfnamefont {Y.-X.}\ \bibnamefont {Wang}},\ and\ \bibinfo {author} {\bibfnamefont {F.}~\bibnamefont {Li}},\ }\bibfield  {title} {\bibinfo {title} {Generic theory of characterizing topological phases under quantum slow dynamics},\ }\href {https://doi.org/10.1103/PhysRevA.106.022219} {\bibfield  {journal} {\bibinfo  {journal} {Phys. Rev. A}\ }\textbf {\bibinfo {volume} {106}},\ \bibinfo {pages} {022219} (\bibinfo {year} {2022})}\BibitemShut {NoStop}%
\bibitem [{\citenamefont {Wu}\ \emph {et~al.}(2023)\citenamefont {Wu}, \citenamefont {Fang},\ and\ \citenamefont {Li}}]{wu_2023}%
  \BibitemOpen
  \bibfield  {author} {\bibinfo {author} {\bibfnamefont {X.}~\bibnamefont {Wu}}, \bibinfo {author} {\bibfnamefont {P.}~\bibnamefont {Fang}},\ and\ \bibinfo {author} {\bibfnamefont {F.}~\bibnamefont {Li}},\ }\bibfield  {title} {\bibinfo {title} {Dynamical characterization of topological phases beyond the minimal models},\ }\href {https://doi.org/10.1103/PhysRevA.107.052209} {\bibfield  {journal} {\bibinfo  {journal} {Phys. Rev. A}\ }\textbf {\bibinfo {volume} {107}},\ \bibinfo {pages} {052209} (\bibinfo {year} {2023})}\BibitemShut {NoStop}%
\bibitem [{\citenamefont {Song}\ and\ \citenamefont {Prodan}(2014)}]{song_2014}%
  \BibitemOpen
  \bibfield  {author} {\bibinfo {author} {\bibfnamefont {J.}~\bibnamefont {Song}}\ and\ \bibinfo {author} {\bibfnamefont {E.}~\bibnamefont {Prodan}},\ }\bibfield  {title} {\bibinfo {title} {Aiii and bdi topological systems at strong disorder},\ }\href {https://doi.org/10.1103/PhysRevB.89.224203} {\bibfield  {journal} {\bibinfo  {journal} {Phys. Rev. B}\ }\textbf {\bibinfo {volume} {89}},\ \bibinfo {pages} {224203} (\bibinfo {year} {2014})}\BibitemShut {NoStop}%
\bibitem [{\citenamefont {Tao}\ \emph {et~al.}(2020)\citenamefont {Tao}, \citenamefont {Yan}, \citenamefont {Liu}, \citenamefont {Niu}, \citenamefont {Zhou}, \citenamefont {Zhang}, \citenamefont {Jia}, \citenamefont {Chen}, \citenamefont {Liu}, \citenamefont {Chen},\ and\ \citenamefont {Yu}}]{tao_2020}%
  \BibitemOpen
  \bibfield  {author} {\bibinfo {author} {\bibfnamefont {Z.}~\bibnamefont {Tao}}, \bibinfo {author} {\bibfnamefont {T.}~\bibnamefont {Yan}}, \bibinfo {author} {\bibfnamefont {W.}~\bibnamefont {Liu}}, \bibinfo {author} {\bibfnamefont {J.}~\bibnamefont {Niu}}, \bibinfo {author} {\bibfnamefont {Y.}~\bibnamefont {Zhou}}, \bibinfo {author} {\bibfnamefont {L.}~\bibnamefont {Zhang}}, \bibinfo {author} {\bibfnamefont {H.}~\bibnamefont {Jia}}, \bibinfo {author} {\bibfnamefont {W.}~\bibnamefont {Chen}}, \bibinfo {author} {\bibfnamefont {S.}~\bibnamefont {Liu}}, \bibinfo {author} {\bibfnamefont {Y.}~\bibnamefont {Chen}},\ and\ \bibinfo {author} {\bibfnamefont {D.}~\bibnamefont {Yu}},\ }\bibfield  {title} {\bibinfo {title} {Simulation of a topological phase transition in a kitaev chain with long-range coupling using a superconducting circuit},\ }\href {https://doi.org/10.1103/PhysRevB.101.035109} {\bibfield  {journal} {\bibinfo  {journal} {Phys. Rev. B}\ }\textbf {\bibinfo {volume} {101}},\ \bibinfo {pages} {035109}
  (\bibinfo {year} {2020})}\BibitemShut {NoStop}%
\bibitem [{\citenamefont {Ran{\v{c}}i{\'{c}}}(2022)}]{rancic_2022}%
  \BibitemOpen
  \bibfield  {author} {\bibinfo {author} {\bibfnamefont {M.~J.}\ \bibnamefont {Ran{\v{c}}i{\'{c}}}},\ }\bibfield  {title} {\bibinfo {title} {Exactly solving the kitaev chain and generating majorana-zero-modes out of noisy qubits},\ }\href {https://doi.org/10.1038/s41598-022-24341-z} {\bibfield  {journal} {\bibinfo  {journal} {Scientific Reports}\ }\textbf {\bibinfo {volume} {12}},\ \bibinfo {pages} {19882} (\bibinfo {year} {2022})}\BibitemShut {NoStop}%
\bibitem [{\citenamefont {Dvir}\ \emph {et~al.}(2023)\citenamefont {Dvir}, \citenamefont {Wang}, \citenamefont {van Loo}, \citenamefont {Liu}, \citenamefont {Mazur}, \citenamefont {Bordin}, \citenamefont {ten Haaf}, \citenamefont {Wang}, \citenamefont {van Driel}, \citenamefont {Zatelli}, \citenamefont {Li}, \citenamefont {Malinowski}, \citenamefont {Gazibegovic}, \citenamefont {Badawy}, \citenamefont {Bakkers}, \citenamefont {Wimmer},\ and\ \citenamefont {Kouwenhoven}}]{dvir_2023}%
  \BibitemOpen
  \bibfield  {author} {\bibinfo {author} {\bibfnamefont {T.}~\bibnamefont {Dvir}}, \bibinfo {author} {\bibfnamefont {G.}~\bibnamefont {Wang}}, \bibinfo {author} {\bibfnamefont {N.}~\bibnamefont {van Loo}}, \bibinfo {author} {\bibfnamefont {C.-X.}\ \bibnamefont {Liu}}, \bibinfo {author} {\bibfnamefont {G.~P.}\ \bibnamefont {Mazur}}, \bibinfo {author} {\bibfnamefont {A.}~\bibnamefont {Bordin}}, \bibinfo {author} {\bibfnamefont {S.~L.~D.}\ \bibnamefont {ten Haaf}}, \bibinfo {author} {\bibfnamefont {J.-Y.}\ \bibnamefont {Wang}}, \bibinfo {author} {\bibfnamefont {D.}~\bibnamefont {van Driel}}, \bibinfo {author} {\bibfnamefont {F.}~\bibnamefont {Zatelli}}, \bibinfo {author} {\bibfnamefont {X.}~\bibnamefont {Li}}, \bibinfo {author} {\bibfnamefont {F.~K.}\ \bibnamefont {Malinowski}}, \bibinfo {author} {\bibfnamefont {S.}~\bibnamefont {Gazibegovic}}, \bibinfo {author} {\bibfnamefont {G.}~\bibnamefont {Badawy}}, \bibinfo {author} {\bibfnamefont {E.~P. A.~M.}\ \bibnamefont {Bakkers}}, \bibinfo {author} {\bibfnamefont
  {M.}~\bibnamefont {Wimmer}},\ and\ \bibinfo {author} {\bibfnamefont {L.~P.}\ \bibnamefont {Kouwenhoven}},\ }\bibfield  {title} {\bibinfo {title} {Realization of a minimal kitaev chain in coupled quantum dots},\ }\href {https://doi.org/10.1038/s41586-022-05585-1} {\bibfield  {journal} {\bibinfo  {journal} {Nature}\ }\textbf {\bibinfo {volume} {614}},\ \bibinfo {pages} {445} (\bibinfo {year} {2023})}\BibitemShut {NoStop}%
\bibitem [{\citenamefont {Kitaev}(2001)}]{kitaev_2001}%
  \BibitemOpen
  \bibfield  {author} {\bibinfo {author} {\bibfnamefont {A.~Y.}\ \bibnamefont {Kitaev}},\ }\bibfield  {title} {\bibinfo {title} {Unpaired majorana fermions in quantum wires},\ }\href {https://doi.org/10.1070/1063-7869/44/10S/S29} {\bibfield  {journal} {\bibinfo  {journal} {Phys.-Usp.}\ }\textbf {\bibinfo {volume} {44}},\ \bibinfo {pages} {131} (\bibinfo {year} {2001})}\BibitemShut {NoStop}%
\bibitem [{\citenamefont {Vidal}\ \emph {et~al.}(2003)\citenamefont {Vidal}, \citenamefont {Latorre}, \citenamefont {Rico},\ and\ \citenamefont {Kitaev}}]{vidal_2003}%
  \BibitemOpen
  \bibfield  {author} {\bibinfo {author} {\bibfnamefont {G.}~\bibnamefont {Vidal}}, \bibinfo {author} {\bibfnamefont {J.~I.}\ \bibnamefont {Latorre}}, \bibinfo {author} {\bibfnamefont {E.}~\bibnamefont {Rico}},\ and\ \bibinfo {author} {\bibfnamefont {A.}~\bibnamefont {Kitaev}},\ }\bibfield  {title} {\bibinfo {title} {Entanglement in quantum critical phenomena},\ }\href {https://doi.org/10.1103/PhysRevLett.90.227902} {\bibfield  {journal} {\bibinfo  {journal} {Phys. Rev. Lett.}\ }\textbf {\bibinfo {volume} {90}},\ \bibinfo {pages} {227902} (\bibinfo {year} {2003})}\BibitemShut {NoStop}%
\bibitem [{\citenamefont {Calabrese}\ and\ \citenamefont {Cardy}(2004)}]{calabrese_2004}%
  \BibitemOpen
  \bibfield  {author} {\bibinfo {author} {\bibfnamefont {P.}~\bibnamefont {Calabrese}}\ and\ \bibinfo {author} {\bibfnamefont {J.}~\bibnamefont {Cardy}},\ }\bibfield  {title} {\bibinfo {title} {Entanglement entropy and quantum field theory},\ }\href {https://doi.org/10.1088/1742-5468/2004/06/P06002} {\bibfield  {journal} {\bibinfo  {journal} {Journal of Statistical Mechanics: Theory and Experiment}\ }\textbf {\bibinfo {volume} {2004}},\ \bibinfo {pages} {P06002} (\bibinfo {year} {2004})}\BibitemShut {NoStop}%
\bibitem [{\citenamefont {Eisert}\ \emph {et~al.}(2010)\citenamefont {Eisert}, \citenamefont {Cramer},\ and\ \citenamefont {Plenio}}]{eisert_2010}%
  \BibitemOpen
  \bibfield  {author} {\bibinfo {author} {\bibfnamefont {J.}~\bibnamefont {Eisert}}, \bibinfo {author} {\bibfnamefont {M.}~\bibnamefont {Cramer}},\ and\ \bibinfo {author} {\bibfnamefont {M.~B.}\ \bibnamefont {Plenio}},\ }\bibfield  {title} {\bibinfo {title} {Colloquium: Area laws for the entanglement entropy},\ }\href {https://doi.org/10.1103/RevModPhys.82.277} {\bibfield  {journal} {\bibinfo  {journal} {Rev. Mod. Phys.}\ }\textbf {\bibinfo {volume} {82}},\ \bibinfo {pages} {277} (\bibinfo {year} {2010})}\BibitemShut {NoStop}%
\bibitem [{\citenamefont {Porras}\ and\ \citenamefont {Cirac}(2004)}]{porras_2004}%
  \BibitemOpen
  \bibfield  {author} {\bibinfo {author} {\bibfnamefont {D.}~\bibnamefont {Porras}}\ and\ \bibinfo {author} {\bibfnamefont {J.~I.}\ \bibnamefont {Cirac}},\ }\bibfield  {title} {\bibinfo {title} {Effective quantum spin systems with trapped ions},\ }\href {https://doi.org/10.1103/PhysRevLett.92.207901} {\bibfield  {journal} {\bibinfo  {journal} {Phys. Rev. Lett.}\ }\textbf {\bibinfo {volume} {92}},\ \bibinfo {pages} {207901} (\bibinfo {year} {2004})}\BibitemShut {NoStop}%
\bibitem [{\citenamefont {Jiang}\ \emph {et~al.}(2011)\citenamefont {Jiang}, \citenamefont {Kitagawa}, \citenamefont {Alicea}, \citenamefont {Akhmerov}, \citenamefont {Pekker}, \citenamefont {Refael}, \citenamefont {Cirac}, \citenamefont {Demler}, \citenamefont {Lukin},\ and\ \citenamefont {Zoller}}]{jiang_2011}%
  \BibitemOpen
  \bibfield  {author} {\bibinfo {author} {\bibfnamefont {L.}~\bibnamefont {Jiang}}, \bibinfo {author} {\bibfnamefont {T.}~\bibnamefont {Kitagawa}}, \bibinfo {author} {\bibfnamefont {J.}~\bibnamefont {Alicea}}, \bibinfo {author} {\bibfnamefont {A.~R.}\ \bibnamefont {Akhmerov}}, \bibinfo {author} {\bibfnamefont {D.}~\bibnamefont {Pekker}}, \bibinfo {author} {\bibfnamefont {G.}~\bibnamefont {Refael}}, \bibinfo {author} {\bibfnamefont {J.~I.}\ \bibnamefont {Cirac}}, \bibinfo {author} {\bibfnamefont {E.}~\bibnamefont {Demler}}, \bibinfo {author} {\bibfnamefont {M.~D.}\ \bibnamefont {Lukin}},\ and\ \bibinfo {author} {\bibfnamefont {P.}~\bibnamefont {Zoller}},\ }\bibfield  {title} {\bibinfo {title} {Majorana fermions in equilibrium and in driven cold-atom quantum wires},\ }\href {https://doi.org/10.1103/PhysRevLett.106.220402} {\bibfield  {journal} {\bibinfo  {journal} {Phys. Rev. Lett.}\ }\textbf {\bibinfo {volume} {106}},\ \bibinfo {pages} {220402} (\bibinfo {year} {2011})}\BibitemShut {NoStop}%
\bibitem [{\citenamefont {Laflamme}\ \emph {et~al.}(2014)\citenamefont {Laflamme}, \citenamefont {Baranov}, \citenamefont {Zoller},\ and\ \citenamefont {Kraus}}]{laflamme_2014}%
  \BibitemOpen
  \bibfield  {author} {\bibinfo {author} {\bibfnamefont {C.}~\bibnamefont {Laflamme}}, \bibinfo {author} {\bibfnamefont {M.~A.}\ \bibnamefont {Baranov}}, \bibinfo {author} {\bibfnamefont {P.}~\bibnamefont {Zoller}},\ and\ \bibinfo {author} {\bibfnamefont {C.~V.}\ \bibnamefont {Kraus}},\ }\bibfield  {title} {\bibinfo {title} {Hybrid topological quantum computation with majorana fermions: A cold-atom setup},\ }\href {https://doi.org/10.1103/PhysRevA.89.022319} {\bibfield  {journal} {\bibinfo  {journal} {Phys. Rev. A}\ }\textbf {\bibinfo {volume} {89}},\ \bibinfo {pages} {022319} (\bibinfo {year} {2014})}\BibitemShut {NoStop}%
\bibitem [{\citenamefont {Anderson}(1958)}]{anderson_1958}%
  \BibitemOpen
  \bibfield  {author} {\bibinfo {author} {\bibfnamefont {P.~W.}\ \bibnamefont {Anderson}},\ }\bibfield  {title} {\bibinfo {title} {Absence of diffusion in certain random lattices},\ }\href {https://doi.org/10.1103/PhysRev.109.1492} {\bibfield  {journal} {\bibinfo  {journal} {Phys. Rev.}\ }\textbf {\bibinfo {volume} {109}},\ \bibinfo {pages} {1492} (\bibinfo {year} {1958})}\BibitemShut {NoStop}%
\bibitem [{\citenamefont {Gurarie}(2011)}]{gurarie_2011}%
  \BibitemOpen
  \bibfield  {author} {\bibinfo {author} {\bibfnamefont {V.}~\bibnamefont {Gurarie}},\ }\bibfield  {title} {\bibinfo {title} {Single-particle green’s functions and interacting topological insulators},\ }\href {https://doi.org/10.1103/PhysRevB.83.085426} {\bibfield  {journal} {\bibinfo  {journal} {Phys. Rev. B}\ }\textbf {\bibinfo {volume} {83}},\ \bibinfo {pages} {085426} (\bibinfo {year} {2011})}\BibitemShut {NoStop}%
\bibitem [{\citenamefont {Manmana}\ \emph {et~al.}(2012)\citenamefont {Manmana}, \citenamefont {Essin}, \citenamefont {Noack},\ and\ \citenamefont {Gurarie}}]{manmana_2012}%
  \BibitemOpen
  \bibfield  {author} {\bibinfo {author} {\bibfnamefont {S.~R.}\ \bibnamefont {Manmana}}, \bibinfo {author} {\bibfnamefont {A.~M.}\ \bibnamefont {Essin}}, \bibinfo {author} {\bibfnamefont {R.~M.}\ \bibnamefont {Noack}},\ and\ \bibinfo {author} {\bibfnamefont {V.}~\bibnamefont {Gurarie}},\ }\bibfield  {title} {\bibinfo {title} {Topological invariants and interacting one-dimensional fermionic systems},\ }\href {https://doi.org/10.1103/PhysRevB.86.205119} {\bibfield  {journal} {\bibinfo  {journal} {Phys. Rev. B}\ }\textbf {\bibinfo {volume} {86}},\ \bibinfo {pages} {205119} (\bibinfo {year} {2012})}\BibitemShut {NoStop}%
\bibitem [{\citenamefont {Chang}(2018)}]{chang_2018}%
  \BibitemOpen
  \bibfield  {author} {\bibinfo {author} {\bibfnamefont {P.-Y.}\ \bibnamefont {Chang}},\ }\bibfield  {title} {\bibinfo {title} {Topology and entanglement in quench dynamics},\ }\href {https://doi.org/10.1103/PhysRevB.97.224304} {\bibfield  {journal} {\bibinfo  {journal} {Phys. Rev. B}\ }\textbf {\bibinfo {volume} {97}},\ \bibinfo {pages} {224304} (\bibinfo {year} {2018})}\BibitemShut {NoStop}%
\bibitem [{\citenamefont {Yang}\ \emph {et~al.}(2018)\citenamefont {Yang}, \citenamefont {Li},\ and\ \citenamefont {Chen}}]{yang_2018}%
  \BibitemOpen
  \bibfield  {author} {\bibinfo {author} {\bibfnamefont {C.}~\bibnamefont {Yang}}, \bibinfo {author} {\bibfnamefont {L.}~\bibnamefont {Li}},\ and\ \bibinfo {author} {\bibfnamefont {S.}~\bibnamefont {Chen}},\ }\bibfield  {title} {\bibinfo {title} {Dynamical topological invariant after a quantum quench},\ }\href {https://link.aps.org/doi/10.1103/PhysRevB.97.060304} {\bibfield  {journal} {\bibinfo  {journal} {Phys. Rev. B}\ }\textbf {\bibinfo {volume} {97}},\ \bibinfo {pages} {060304(R)} (\bibinfo {year} {2018})}\BibitemShut {NoStop}%
\bibitem [{\citenamefont {Hsu}\ \emph {et~al.}(2021)\citenamefont {Hsu}, \citenamefont {Chiu},\ and\ \citenamefont {Chang}}]{hsu_2021}%
  \BibitemOpen
  \bibfield  {author} {\bibinfo {author} {\bibfnamefont {H.-C.}\ \bibnamefont {Hsu}}, \bibinfo {author} {\bibfnamefont {P.-M.}\ \bibnamefont {Chiu}},\ and\ \bibinfo {author} {\bibfnamefont {P.-Y.}\ \bibnamefont {Chang}},\ }\bibfield  {title} {\bibinfo {title} {Disorder-induced topology in quench dynamics},\ }\href {https://link.aps.org/doi/10.1103/PhysRevResearch.3.033242} {\bibfield  {journal} {\bibinfo  {journal} {Phys. Rev. Res.}\ }\textbf {\bibinfo {volume} {3}},\ \bibinfo {pages} {033242} (\bibinfo {year} {2021})}\BibitemShut {NoStop}%
\bibitem [{\citenamefont {Zak}(1989)}]{zak_1989}%
  \BibitemOpen
  \bibfield  {author} {\bibinfo {author} {\bibfnamefont {J.}~\bibnamefont {Zak}},\ }\bibfield  {title} {\bibinfo {title} {Berry's phase for energy bands in solids},\ }\href {https://doi.org/10.1103/PhysRevLett.62.2747} {\bibfield  {journal} {\bibinfo  {journal} {Phys. Rev. Lett.}\ }\textbf {\bibinfo {volume} {62}},\ \bibinfo {pages} {2747} (\bibinfo {year} {1989})}\BibitemShut {NoStop}%
\bibitem [{\citenamefont {Budich}\ and\ \citenamefont {Ardonne}(2013)}]{budich_2013}%
  \BibitemOpen
  \bibfield  {author} {\bibinfo {author} {\bibfnamefont {J.~C.}\ \bibnamefont {Budich}}\ and\ \bibinfo {author} {\bibfnamefont {E.}~\bibnamefont {Ardonne}},\ }\bibfield  {title} {\bibinfo {title} {Equivalent topological invariants for one-dimensional majorana wires in symmetry class d},\ }\href {https://doi.org/10.1103/PhysRevB.88.075419} {\bibfield  {journal} {\bibinfo  {journal} {Phys. Rev. B}\ }\textbf {\bibinfo {volume} {88}},\ \bibinfo {pages} {075419} (\bibinfo {year} {2013})}\BibitemShut {NoStop}%
\bibitem [{\citenamefont {Rahul}\ \emph {et~al.}(2019)\citenamefont {Rahul}, \citenamefont {Kumar}, \citenamefont {Kartik}, \citenamefont {Banerjee},\ and\ \citenamefont {Sarkar}}]{rahul_2019}%
  \BibitemOpen
  \bibfield  {author} {\bibinfo {author} {\bibfnamefont {S.}~\bibnamefont {Rahul}}, \bibinfo {author} {\bibfnamefont {R.~R.}\ \bibnamefont {Kumar}}, \bibinfo {author} {\bibfnamefont {Y.~R.}\ \bibnamefont {Kartik}}, \bibinfo {author} {\bibfnamefont {A.}~\bibnamefont {Banerjee}},\ and\ \bibinfo {author} {\bibfnamefont {S.}~\bibnamefont {Sarkar}},\ }\bibfield  {title} {\bibinfo {title} {An interplay of topology and quantized geometric phase for two different symmetry-class hamiltonians},\ }\href {https://doi.org/10.1088/1402-4896/ab1d7b} {\bibfield  {journal} {\bibinfo  {journal} {Phys. Scr.}\ }\textbf {\bibinfo {volume} {94}},\ \bibinfo {pages} {115803} (\bibinfo {year} {2019})}\BibitemShut {NoStop}%
\bibitem [{Note1()}]{Note1}%
  \BibitemOpen
  \bibinfo {note} {While for an individual realization the left- and right-localized MZMs will, in general, have different profiles, on average we find that they qualitatively exhibit the same dynamics. As such, we choose to discuss solely the left-localized modes for brevity.}\BibitemShut {Stop}%
\bibitem [{\citenamefont {Auckly}(1994)}]{auckly_1994}%
  \BibitemOpen
  \bibfield  {author} {\bibinfo {author} {\bibfnamefont {D.~R.}\ \bibnamefont {Auckly}},\ }\bibfield  {title} {\bibinfo {title} {Topological methods to compute chern-simons invariants},\ }\href {https://doi.org/10.1017/S0305004100072066} {\bibfield  {journal} {\bibinfo  {journal} {Math. Proc. Cambridge Philos. Soc.}\ }\textbf {\bibinfo {volume} {115}},\ \bibinfo {pages} {229–251} (\bibinfo {year} {1994})}\BibitemShut {NoStop}%
\bibitem [{\citenamefont {Fisher}(1965)}]{fisher_1965}%
  \BibitemOpen
  \bibfield  {author} {\bibinfo {author} {\bibfnamefont {M.~E.}\ \bibnamefont {Fisher}},\ }\href@noop {} {\emph {\bibinfo {title} {The nature of critical points}}}\ (\bibinfo  {publisher} {University of Colorado Press},\ \bibinfo {year} {1965})\BibitemShut {NoStop}%
\bibitem [{\citenamefont {Fisher}(1967)}]{fisher_1967}%
  \BibitemOpen
  \bibfield  {author} {\bibinfo {author} {\bibfnamefont {M.~E.}\ \bibnamefont {Fisher}},\ }\bibfield  {title} {\bibinfo {title} {The theory of equilibrium critical phenomena},\ }\href {https://doi.org/10.1088/0034-4885/30/2/306} {\bibfield  {journal} {\bibinfo  {journal} {Rep. Prog. Phys.}\ }\textbf {\bibinfo {volume} {30}},\ \bibinfo {pages} {615} (\bibinfo {year} {1967})}\BibitemShut {NoStop}%
\bibitem [{\citenamefont {Jalabert}\ and\ \citenamefont {Pastawski}(2001)}]{jalabert_2001}%
  \BibitemOpen
  \bibfield  {author} {\bibinfo {author} {\bibfnamefont {R.~A.}\ \bibnamefont {Jalabert}}\ and\ \bibinfo {author} {\bibfnamefont {H.~M.}\ \bibnamefont {Pastawski}},\ }\bibfield  {title} {\bibinfo {title} {Environment-independent decoherence rate in classically chaotic systems},\ }\href {https://doi.org/10.1103/PhysRevLett.86.2490} {\bibfield  {journal} {\bibinfo  {journal} {Phys. Rev. Lett.}\ }\textbf {\bibinfo {volume} {86}},\ \bibinfo {pages} {2490} (\bibinfo {year} {2001})}\BibitemShut {NoStop}%
\bibitem [{\citenamefont {Vajna}\ and\ \citenamefont {D\'ora}(2015)}]{vajna_2015}%
  \BibitemOpen
  \bibfield  {author} {\bibinfo {author} {\bibfnamefont {S.}~\bibnamefont {Vajna}}\ and\ \bibinfo {author} {\bibfnamefont {B.}~\bibnamefont {D\'ora}},\ }\bibfield  {title} {\bibinfo {title} {Topological classification of dynamical phase transitions},\ }\href {https://doi.org/10.1103/PhysRevB.91.155127} {\bibfield  {journal} {\bibinfo  {journal} {Phys. Rev. B}\ }\textbf {\bibinfo {volume} {91}},\ \bibinfo {pages} {155127} (\bibinfo {year} {2015})}\BibitemShut {NoStop}%
\bibitem [{\citenamefont {Budich}\ and\ \citenamefont {Heyl}(2016)}]{budich_2016}%
  \BibitemOpen
  \bibfield  {author} {\bibinfo {author} {\bibfnamefont {J.~C.}\ \bibnamefont {Budich}}\ and\ \bibinfo {author} {\bibfnamefont {M.}~\bibnamefont {Heyl}},\ }\bibfield  {title} {\bibinfo {title} {Dynamical topological order parameters far from equilibrium},\ }\href {https://doi.org/10.1103/PhysRevB.93.085416} {\bibfield  {journal} {\bibinfo  {journal} {Phys. Rev. B}\ }\textbf {\bibinfo {volume} {93}},\ \bibinfo {pages} {085416} (\bibinfo {year} {2016})}\BibitemShut {NoStop}%
\bibitem [{\citenamefont {Zhang}\ and\ \citenamefont {Song}(2020)}]{zhang_2020}%
  \BibitemOpen
  \bibfield  {author} {\bibinfo {author} {\bibfnamefont {K.~L.}\ \bibnamefont {Zhang}}\ and\ \bibinfo {author} {\bibfnamefont {Z.}~\bibnamefont {Song}},\ }\bibfield  {title} {\bibinfo {title} {Manifestation of a topological gapless phase in a two-dimensional chiral symmetric system through loschmidt echo},\ }\href {https://doi.org/10.1103/PhysRevB.101.014303} {\bibfield  {journal} {\bibinfo  {journal} {Phys. Rev. B}\ }\textbf {\bibinfo {volume} {101}},\ \bibinfo {pages} {014303} (\bibinfo {year} {2020})}\BibitemShut {NoStop}%
\bibitem [{\citenamefont {Anderson}(1967)}]{anderson_1967}%
  \BibitemOpen
  \bibfield  {author} {\bibinfo {author} {\bibfnamefont {P.~W.}\ \bibnamefont {Anderson}},\ }\bibfield  {title} {\bibinfo {title} {Infrared catastrophe in fermi gases with local scattering potentials},\ }\href {https://doi.org/10.1103/PhysRevLett.18.1049} {\bibfield  {journal} {\bibinfo  {journal} {Phys. Rev. Lett.}\ }\textbf {\bibinfo {volume} {18}},\ \bibinfo {pages} {1049} (\bibinfo {year} {1967})}\BibitemShut {NoStop}%
\bibitem [{\citenamefont {Sedlmayr}(2019)}]{sedlmayr_2019}%
  \BibitemOpen
  \bibfield  {author} {\bibinfo {author} {\bibfnamefont {N.}~\bibnamefont {Sedlmayr}},\ }\bibfield  {title} {\bibinfo {title} {Dynamical phase transitions in topological insulators},\ }\href {https://doi.org/10.12693/APhysPolA.135.1191} {\bibfield  {journal} {\bibinfo  {journal} {Acta Phys. Pol. A}\ }\textbf {\bibinfo {volume} {135}},\ \bibinfo {pages} {1191} (\bibinfo {year} {2019})}\BibitemShut {NoStop}%
\bibitem [{\citenamefont {Heyl}(2018)}]{heyl_2018}%
  \BibitemOpen
  \bibfield  {author} {\bibinfo {author} {\bibfnamefont {M.}~\bibnamefont {Heyl}},\ }\bibfield  {title} {\bibinfo {title} {Dynamical quantum phase transitions: a review},\ }\href {https://doi.org/10.1088/1361-6633/aaaf9a} {\bibfield  {journal} {\bibinfo  {journal} {Rep. Prog. Phys.}\ }\textbf {\bibinfo {volume} {81}},\ \bibinfo {pages} {054001} (\bibinfo {year} {2018})}\BibitemShut {NoStop}%
\bibitem [{\citenamefont {Vanhala}\ and\ \citenamefont {Ojanen}(2023)}]{vanhala_2023}%
  \BibitemOpen
  \bibfield  {author} {\bibinfo {author} {\bibfnamefont {T.~I.}\ \bibnamefont {Vanhala}}\ and\ \bibinfo {author} {\bibfnamefont {T.}~\bibnamefont {Ojanen}},\ }\bibfield  {title} {\bibinfo {title} {Theory of the loschmidt echo and dynamical quantum phase transitions in disordered fermi systems},\ }\href {https://doi.org/10.1103/PhysRevResearch.5.033178} {\bibfield  {journal} {\bibinfo  {journal} {Phys. Rev. Res.}\ }\textbf {\bibinfo {volume} {5}},\ \bibinfo {pages} {033178} (\bibinfo {year} {2023})}\BibitemShut {NoStop}%
\bibitem [{\citenamefont {Sedlmayr}\ and\ \citenamefont {Bena}(2015)}]{sedlmayr_2015}%
  \BibitemOpen
  \bibfield  {author} {\bibinfo {author} {\bibfnamefont {N.}~\bibnamefont {Sedlmayr}}\ and\ \bibinfo {author} {\bibfnamefont {C.}~\bibnamefont {Bena}},\ }\bibfield  {title} {\bibinfo {title} {Visualizing majorana bound states in one and two dimensions using the generalized majorana polarization},\ }\href {https://doi.org/10.1103/PhysRevB.92.115115} {\bibfield  {journal} {\bibinfo  {journal} {Phys. Rev. B}\ }\textbf {\bibinfo {volume} {92}},\ \bibinfo {pages} {115115} (\bibinfo {year} {2015})}\BibitemShut {NoStop}%
\bibitem [{Note2()}]{Note2}%
  \BibitemOpen
  \bibinfo {note} {This is a direct consequence of the initial discrete symmetries: $(i)$ TRS makes $\protect \mathcal {H}(t=0)$ real when $\Delta $ is real; $(ii)$ PHS maps any finite energy state into an orthogonal one, except for the Majoranas. The ChS is a composition of TRS and PHS. TRS is represented by complex conjugation because of $(i)$, so the corresponding PHS eigenvalues $\pm 1$ classify the MZMs}\BibitemShut {NoStop}%
\bibitem [{\citenamefont {Li}\ and\ \citenamefont {Haldane}(2008)}]{li_2008}%
  \BibitemOpen
  \bibfield  {author} {\bibinfo {author} {\bibfnamefont {H.}~\bibnamefont {Li}}\ and\ \bibinfo {author} {\bibfnamefont {F.~D.~M.}\ \bibnamefont {Haldane}},\ }\bibfield  {title} {\bibinfo {title} {Entanglement spectrum as a generalization of entanglement entropy: Identification of topological order in non-abelian fractional quantum hall effect states},\ }\href {https://doi.org/10.1103/PhysRevLett.101.010504} {\bibfield  {journal} {\bibinfo  {journal} {Phys. Rev. Lett.}\ }\textbf {\bibinfo {volume} {101}},\ \bibinfo {pages} {010504} (\bibinfo {year} {2008})}\BibitemShut {NoStop}%
\bibitem [{\citenamefont {Legner}\ and\ \citenamefont {Neupert}(2013)}]{legner_2013}%
  \BibitemOpen
  \bibfield  {author} {\bibinfo {author} {\bibfnamefont {M.}~\bibnamefont {Legner}}\ and\ \bibinfo {author} {\bibfnamefont {T.}~\bibnamefont {Neupert}},\ }\bibfield  {title} {\bibinfo {title} {Relating the entanglement spectrum of noninteracting band insulators to their quantum geometry and topology},\ }\href {https://doi.org/10.1103/PhysRevB.88.115114} {\bibfield  {journal} {\bibinfo  {journal} {Phys. Rev. B}\ }\textbf {\bibinfo {volume} {88}},\ \bibinfo {pages} {115114} (\bibinfo {year} {2013})}\BibitemShut {NoStop}%
\bibitem [{\citenamefont {Patrick}\ \emph {et~al.}(2017)\citenamefont {Patrick}, \citenamefont {Neupert},\ and\ \citenamefont {Pachos}}]{patrick_2017}%
  \BibitemOpen
  \bibfield  {author} {\bibinfo {author} {\bibfnamefont {K.}~\bibnamefont {Patrick}}, \bibinfo {author} {\bibfnamefont {T.}~\bibnamefont {Neupert}},\ and\ \bibinfo {author} {\bibfnamefont {J.~K.}\ \bibnamefont {Pachos}},\ }\bibfield  {title} {\bibinfo {title} {Topological quantum liquids with long-range couplings},\ }\href {https://doi.org/10.1103/PhysRevLett.118.267002} {\bibfield  {journal} {\bibinfo  {journal} {Phys. Rev. Lett.}\ }\textbf {\bibinfo {volume} {118}},\ \bibinfo {pages} {267002} (\bibinfo {year} {2017})}\BibitemShut {NoStop}%
\bibitem [{\citenamefont {Fidkowski}(2010)}]{fidkowski_2010}%
  \BibitemOpen
  \bibfield  {author} {\bibinfo {author} {\bibfnamefont {L.}~\bibnamefont {Fidkowski}},\ }\bibfield  {title} {\bibinfo {title} {Entanglement spectrum of topological insulators and superconductors},\ }\href {https://doi.org/10.1103/PhysRevLett.104.130502} {\bibfield  {journal} {\bibinfo  {journal} {Phys. Rev. Lett.}\ }\textbf {\bibinfo {volume} {104}},\ \bibinfo {pages} {130502} (\bibinfo {year} {2010})}\BibitemShut {NoStop}%
\bibitem [{\citenamefont {Chandran}\ \emph {et~al.}(2011)\citenamefont {Chandran}, \citenamefont {Hermanns}, \citenamefont {Regnault},\ and\ \citenamefont {Bernevig}}]{chandran_2011}%
  \BibitemOpen
  \bibfield  {author} {\bibinfo {author} {\bibfnamefont {A.}~\bibnamefont {Chandran}}, \bibinfo {author} {\bibfnamefont {M.}~\bibnamefont {Hermanns}}, \bibinfo {author} {\bibfnamefont {N.}~\bibnamefont {Regnault}},\ and\ \bibinfo {author} {\bibfnamefont {B.~A.}\ \bibnamefont {Bernevig}},\ }\bibfield  {title} {\bibinfo {title} {Bulk-edge correspondence in entanglement spectra},\ }\href {https://doi.org/10.1103/PhysRevB.84.205136} {\bibfield  {journal} {\bibinfo  {journal} {Phys. Rev. B}\ }\textbf {\bibinfo {volume} {84}},\ \bibinfo {pages} {205136} (\bibinfo {year} {2011})}\BibitemShut {NoStop}%
\bibitem [{\citenamefont {Pollmann}\ \emph {et~al.}(2010)\citenamefont {Pollmann}, \citenamefont {Turner}, \citenamefont {Berg},\ and\ \citenamefont {Oshikawa}}]{pollmann_2010}%
  \BibitemOpen
  \bibfield  {author} {\bibinfo {author} {\bibfnamefont {F.}~\bibnamefont {Pollmann}}, \bibinfo {author} {\bibfnamefont {A.~M.}\ \bibnamefont {Turner}}, \bibinfo {author} {\bibfnamefont {E.}~\bibnamefont {Berg}},\ and\ \bibinfo {author} {\bibfnamefont {M.}~\bibnamefont {Oshikawa}},\ }\bibfield  {title} {\bibinfo {title} {Entanglement spectrum of a topological phase in one dimension},\ }\href {https://doi.org/10.1103/PhysRevB.81.064439} {\bibfield  {journal} {\bibinfo  {journal} {Phys. Rev. B}\ }\textbf {\bibinfo {volume} {81}},\ \bibinfo {pages} {064439} (\bibinfo {year} {2010})}\BibitemShut {NoStop}%
\bibitem [{\citenamefont {Cho}\ and\ \citenamefont {Kim}(2017)}]{cho_2017}%
  \BibitemOpen
  \bibfield  {author} {\bibinfo {author} {\bibfnamefont {J.}~\bibnamefont {Cho}}\ and\ \bibinfo {author} {\bibfnamefont {K.~W.}\ \bibnamefont {Kim}},\ }\bibfield  {title} {\bibinfo {title} {Quantum phase transition and entanglement in topological quantum wires},\ }\href {https://doi.org/10.1038/s41598-017-02717-w} {\bibfield  {journal} {\bibinfo  {journal} {Sci. Rep.}\ }\textbf {\bibinfo {volume} {7}},\ \bibinfo {pages} {2745} (\bibinfo {year} {2017})}\BibitemShut {NoStop}%
\bibitem [{\citenamefont {Peschel}(2003)}]{peschel_2003}%
  \BibitemOpen
  \bibfield  {author} {\bibinfo {author} {\bibfnamefont {I.}~\bibnamefont {Peschel}},\ }\bibfield  {title} {\bibinfo {title} {Calculation of reduced density matrices from correlation functions},\ }\href {https://doi.org/10.1088/0305-4470/36/14/101} {\bibfield  {journal} {\bibinfo  {journal} {J. Phys. A: Math. Gen.}\ }\textbf {\bibinfo {volume} {36}},\ \bibinfo {pages} {L205} (\bibinfo {year} {2003})}\BibitemShut {NoStop}%
\bibitem [{\citenamefont {Peschel}\ and\ \citenamefont {Eisler}(2009)}]{peschel_2009}%
  \BibitemOpen
  \bibfield  {author} {\bibinfo {author} {\bibfnamefont {I.}~\bibnamefont {Peschel}}\ and\ \bibinfo {author} {\bibfnamefont {V.}~\bibnamefont {Eisler}},\ }\bibfield  {title} {\bibinfo {title} {Reduced density matrices and entanglement entropy in free lattice models},\ }\href {https://doi.org/10.1088/1751-8113/42/50/504003} {\bibfield  {journal} {\bibinfo  {journal} {J. Phys. A Math. Theor.}\ }\textbf {\bibinfo {volume} {42}},\ \bibinfo {pages} {504003} (\bibinfo {year} {2009})}\BibitemShut {NoStop}%
\bibitem [{\citenamefont {Senthil}(2015)}]{senthil_2015}%
  \BibitemOpen
  \bibfield  {author} {\bibinfo {author} {\bibfnamefont {T.}~\bibnamefont {Senthil}},\ }\bibfield  {title} {\bibinfo {title} {Symmetry-protected topological phases of quantum matter},\ }\href {https://doi.org/10.1146/annurev-conmatphys-031214-014740} {\bibfield  {journal} {\bibinfo  {journal} {Annu. Rev. Condens. Matter Phys.}\ }\textbf {\bibinfo {volume} {6}},\ \bibinfo {pages} {299} (\bibinfo {year} {2015})}\BibitemShut {NoStop}%
\bibitem [{\citenamefont {Hastings}(2007)}]{hastings_2007}%
  \BibitemOpen
  \bibfield  {author} {\bibinfo {author} {\bibfnamefont {M.~B.}\ \bibnamefont {Hastings}},\ }\bibfield  {title} {\bibinfo {title} {An area law for one-dimensional quantum systems},\ }\href {https://doi.org/10.1088/1742-5468/2007/08/P08024} {\bibfield  {journal} {\bibinfo  {journal} {J. Stat. Mech.}\ }\textbf {\bibinfo {volume} {2007}},\ \bibinfo {pages} {P08024} (\bibinfo {year} {2007})}\BibitemShut {NoStop}%
\bibitem [{\citenamefont {Brand{\~a}o}\ and\ \citenamefont {Horodecki}(2013)}]{brandao_2013}%
  \BibitemOpen
  \bibfield  {author} {\bibinfo {author} {\bibfnamefont {F.~G. S.~L.}\ \bibnamefont {Brand{\~a}o}}\ and\ \bibinfo {author} {\bibfnamefont {M.}~\bibnamefont {Horodecki}},\ }\bibfield  {title} {\bibinfo {title} {An area law for entanglement from exponential decay of correlations},\ }\href {https://doi.org/10.1038/nphys2747} {\bibfield  {journal} {\bibinfo  {journal} {Nat. Phys.}\ }\textbf {\bibinfo {volume} {9}},\ \bibinfo {pages} {721} (\bibinfo {year} {2013})}\BibitemShut {NoStop}%
\bibitem [{\citenamefont {Verresen}\ \emph {et~al.}(2017)\citenamefont {Verresen}, \citenamefont {Moessner},\ and\ \citenamefont {Pollmann}}]{verresen_2017}%
  \BibitemOpen
  \bibfield  {author} {\bibinfo {author} {\bibfnamefont {R.}~\bibnamefont {Verresen}}, \bibinfo {author} {\bibfnamefont {R.}~\bibnamefont {Moessner}},\ and\ \bibinfo {author} {\bibfnamefont {F.}~\bibnamefont {Pollmann}},\ }\bibfield  {title} {\bibinfo {title} {One-dimensional symmetry protected topological phases and their transitions},\ }\href {https://doi.org/10.1103/PhysRevB.96.165124} {\bibfield  {journal} {\bibinfo  {journal} {Phys. Rev. B}\ }\textbf {\bibinfo {volume} {96}},\ \bibinfo {pages} {165124} (\bibinfo {year} {2017})}\BibitemShut {NoStop}%
\end{thebibliography}%

\end{document}